\documentclass{article}

 \usepackage[preprint]{neurips_2026}

\usepackage[utf8]{inputenc} 
\usepackage[T1]{fontenc}    
\usepackage[hidelinks]{hyperref}       
\usepackage{url}            
\usepackage{booktabs}       
\usepackage{amsfonts}       
\usepackage{nicefrac}       
\usepackage{microtype}      
\usepackage{xcolor}         
\usepackage{subcaption}
\usepackage{graphicx}
\usepackage{pifont}
\usepackage{multirow} 
\usepackage{wrapfig}
\usepackage{amsmath}
\usepackage{cleveref}
\usepackage{pgffor}
\usepackage{colortbl}

\title{ReaLB: Real-Time Load Balancing for Multimodal MoE Inference}

\author{%
Yingping Wang, Yi Wu, Xiangyu Wu, Junwei Cui, Weilin Cai, Zhijiang Guo, Jiayi Huang \\
The Hong Kong University of Science and Technology (Guangzhou), China \\
\texttt{hjy@hkust-gz.edu.cn} \thanks{Corresponding author}
}

\begin{document}

\maketitle

\begin{abstract}
\label{sec:abs}
Mixture-of-Experts (MoE) architectures are widely used in modern large language models and multimodal models. 
However, inference efficiency is often limited by highly dynamic and skewed expert workloads across different modalities. During the prefill stage with large batch sizes, vision tokens frequently dominate the input sequences. Under expert parallelism (EP), this leads to severe load imbalance, where a subset of devices becomes overloaded, reducing overall system throughput.
We propose ReaLB, a real-time load balancing method for multimodal MoE (MMoE) inference that introduces zero scheduling overhead. 
ReaLB dynamically adjusts the computation precision of MoE experts at runtime on a per-EP-rank basis. For ranks dominated by vision-heavy experts, ReaLB assigns lower-precision computation to improve execution efficiency by exploiting FP4 Tensor Cores. 
ReaLB does not require redundant experts or additional memory allocation. 
Instead, it performs layer-wise expert precision transformation on the fly and hides the associated overhead within the dispatch phase before MoE computation.
Experiments on representative MMoE models show that ReaLB achieves 1.10$\times$--1.32$\times$ end-to-end speedup while limiting average accuracy degradation to within 1\%.

\end{abstract}

\section{Introduction}
\label{sec:intro}

Mixture-of-Experts (MoE) models are widely adopted in modern generative models~\cite{dsv3,dsvl2}.
From text generation to vision language tasks, MoE supports parameter-efficient scaling by activating only a subset of experts per token.
However, sparse activation and input-dependent routing naturally lead to load imbalance across experts\cite{switch}.
Under expert parallelism (EP), skewed expert selection causes significant inefficiency in both training and inference.
This problem is more severe in multimodal MoE (MMoE) inference~\cite{kimi,qwen3-vl}.

Most prior work focuses on mitigating load imbalance during MoE training.
At the model level, auxiliary balancing losses~\cite{switch,dsv1} and device-constrained routing~\cite{dsv2} regulate gating behavior to improve expert utilization.
These methods, however, are not directly applicable to inference~\cite{dsv3}.
At the system level, FasterMoE~\cite{fastermoe} introduces shadow experts that replicate overloaded experts across devices, at the cost of extra memory usage and communication.
EPLB-like methods~\cite{eplb,384} extend similar ideas to inference, but relies on prediction-based scheduling, which adapts poorly to rapidly changing routing patterns.

\begin{figure}[htbp]
    \centering
    \begin{subfigure}{0.49\linewidth}
        \centering
        \includegraphics[width=\linewidth]{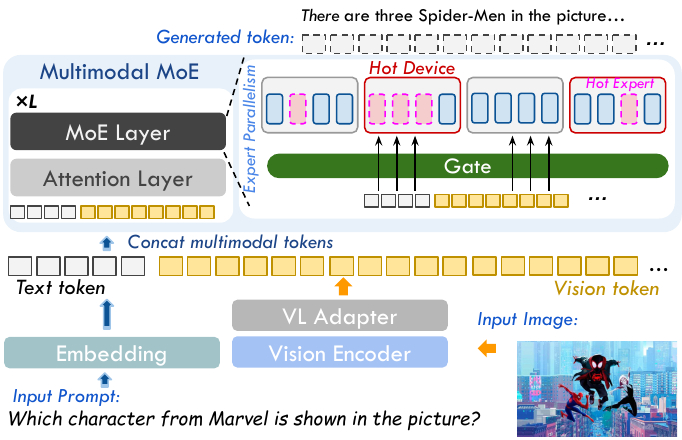}
        \caption{Text-Image to text generation workflow of an MMoE model.}
        \label{fig:intro_1}
    \end{subfigure}
    \hfill
    \begin{subfigure}{0.49\linewidth}
        \centering
        \includegraphics[width=\linewidth]{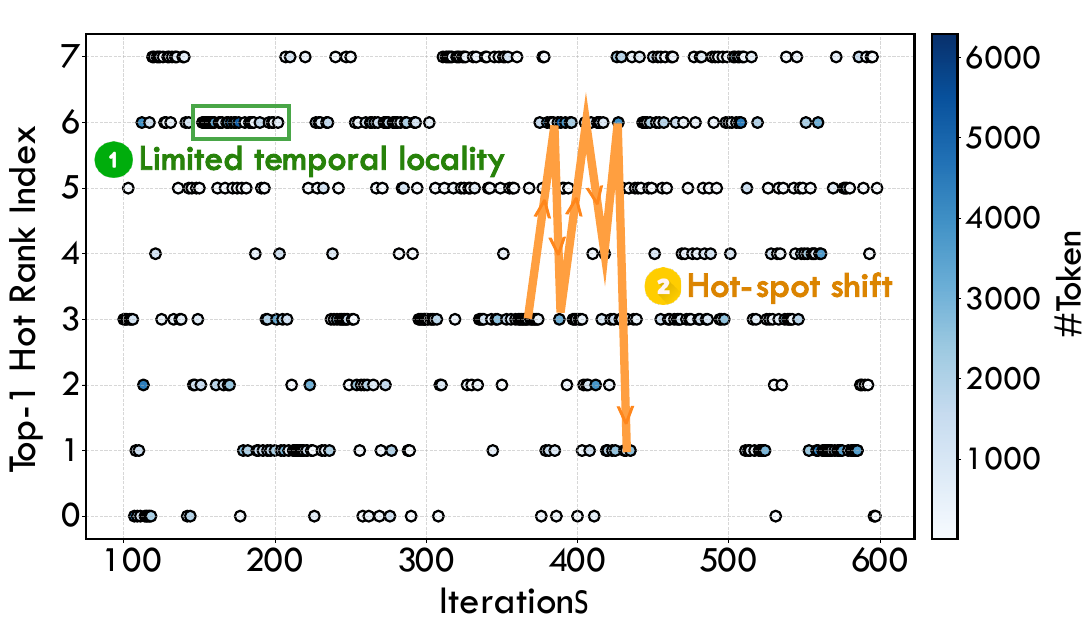}
        \caption{Dynamics of expert selection across eight devices over inference iterations.}
        \label{fig:intro_2}
    \end{subfigure}
    \caption{Motivation for real-time load balancing.}
    \label{fig:intro}
\end{figure}

\Cref{fig:intro} shows the highly dynamic nature of expert selection during MMoE inference~\cite{kimi}.
Unlike the strong temporal locality observed in training~\cite{fastermoe}, expert activation frequencies in inference fluctuate significantly across iterations.
This behavior makes \textbf{stragglers transient and difficult to predict}.
As a result, an effective load balancing strategy must satisfy two stringent requirements:
\textbf{\textit{\ding{172} reacting to fast-changing routing dynamics in real time;}
\textit{\ding{173} incurring minimal additional overhead to remain effective in practical deployment}.}

To meet these requirements, we propose \textbf{ReaLB}, a lightweight, real-time load balancing strategy for large-scale MMoE inference.
Our design is motivated by the observation that, compared to text tokens, 
vision tokens exhibit higher redundancy along the forward pass~\cite{fastv,PyramidDrop}.
We further observe that EP stragglers are often dominated by vision-heavy experts (\Cref{sec:moti}).
ReaLB dynamically balances workloads by switching such experts to lower-precision computation at runtime, reducing their execution latency while preserving accuracy for text tokens.
Instead of storing pre-converted weights, ReaLB performs MoE precision switching on the fly while keeping original high-precision weights intact.
The lightweight transformation cost is fully hidden through pipeline orchestration in the execution flow.
As a result, ReaLB achieves device-level load balance without additional memory footprint or latency overhead.

We implement ReaLB on top of vLLM~\cite{vllm} and evaluate it on Kimi-VL-A3B-Instruct~~\cite{kimi} (Kimi-VL) and Qwen3-VL-30B-A3B-Instruct~\cite{qwen3-vl} (Qwen3-VL), using an 8$\times$ RTX 5090 cluster with EP.
Results show that ReaLB improves end-to-end throughput by up to 1.32$\times$, while maintaining accuracy comparable to BF16 baselines.

Our main contributions are summarized as follows:
\begin{itemize}
\itemsep0em
\item We identify and quantify real-time load imbalance in multimodal MoE inference, highlighting its modality-driven, highly dynamic, and hard-to-predict behavior.
\item We propose \textbf{ReaLB}, a modality-aware, precision-adaptive MoE scheduling mechanism with AIMD-based control for inference-time load balancing.
\item We design a overhead-aware pipeline orchestration method that overlaps online precision switching with all-to-all dispatch, hiding transformation latency from the critical path.
\item We implement ReaLB in vLLM and demonstrate consistent performance gains on large multimodal MoE models with negligible accuracy degradation.
\end{itemize}

\section{Background \& Related Work}
\label{sec:bg}


\subsection{Multimodal MoE}
Multimodal large language models (MLLMs) integrate text and vision processing within a unified architecture.
Compared to dense feed-forward layers, MoE layers provide higher capacity and better scalability for multimodal workloads.
Most existing MMoE architectures combine a Vision Transformer (ViT), a vision-language adapter, and an MoE-augmented LLM backbone.
In this work, we focus on \textit{modality-fused MMoE} designs~\cite{kimi,qwen3-vl,dsv1,internvl3.5,seed1.5vl}, which are widely adopted in recent MLLMs.
In such designs, tokens from different modalities are processed jointly by shared MoE layers, where a unified gating module routes all tokens to a common pool of experts without explicit modality separation.
Recent studies explore modality-isolated experts to reduce cross-modal interference~\cite{cogvlm,eve,moma,ernie}, and these designs are still evolving.

\subsection{MoE Load Balancing Systems}

Sparse activation and input-dependent routing make load imbalance inherent in distributed MoE inference under EP.
In practice, imbalance arises across multiple system dimensions and directly limits throughput.
We summarize two main sources of load imbalance in EP-based MoE systems.

\textbf{Computation.}
Routing decisions in MoE are highly skewed across experts and vary across inputs and modalities.
With static expert-to-GPU mapping under EP, this skew translates into uneven compute load.
The problem is amplified in extremely sparse settings where each GPU hosts a single expert.
During inference, especially in large-batch prefill for multimodal models, vision tokens often dominate routing, creating persistent stragglers.
Most existing load balancing methods target training-time optimization~\cite{smartermoe, megascalemoe, lazarus}, such as auxiliary routing losses~\cite{dsv1} or capacity-based token dropping~\cite{switch}, and are not suitable for inference.
Runtime approaches based on expert replication or replacement~\cite{fastermoe, flexmoe, eplb, inefficiencies, xLLM} attempt to rebalance load, but rely on prediction-based scheduling and introduce extra communication due to expert weight migration.
Recent work, LPLB~\cite{lplb}, extends EPLB and further leverages linear programming (LP) to redistribute token traffic more evenly across experts. 
However, when invoked at high frequency to approximate real-time load balancing, its additional scheduling overhead can outweigh the gains from improved balance, potentially leading to performance degradation.

\textbf{Communication.}
EP requires all-to-all communication for token dispatch and result combination.
Compute imbalance leads to synchronization stalls, where faster GPUs wait for straggling ranks.
Prior work reduces communication cost through overlap and kernel fusion~\cite{scmoe, chimera}, but these techniques assume balanced computation.

\section{Motivation and Problem Formulation}
\label{sec:moti}

\subsection{High Dynamics of MMoE Routing}


We study routing dynamics during MMoE inference by collecting layer-wise routing statistics from Kimi-VL, running on 8 GPUs and profiled with MMMU~\cite{mmmu} over multiple iterations. Here, each iteration corresponds to a single model forward pass under a colocated deployment setting, where prefill and decode are not separated into distinct instances.

\begin{figure}[t]
    \centering
    \begin{subfigure}{0.31\linewidth}
        \centering
        \includegraphics[width=\linewidth]{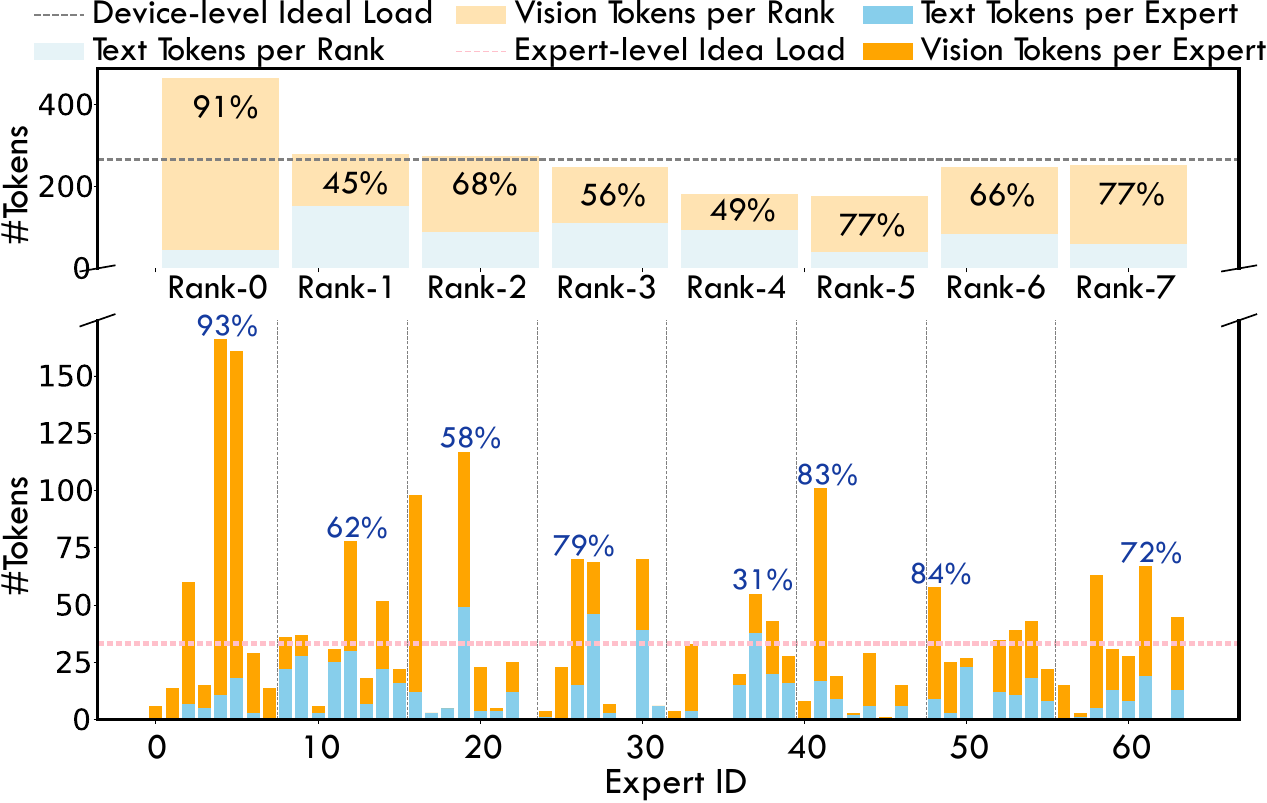}
        \caption{}
        \label{fig:device-dynamics}
    \end{subfigure}
    \hfill
    \begin{subfigure}{0.32\linewidth}
        \centering
        \includegraphics[width=\linewidth]{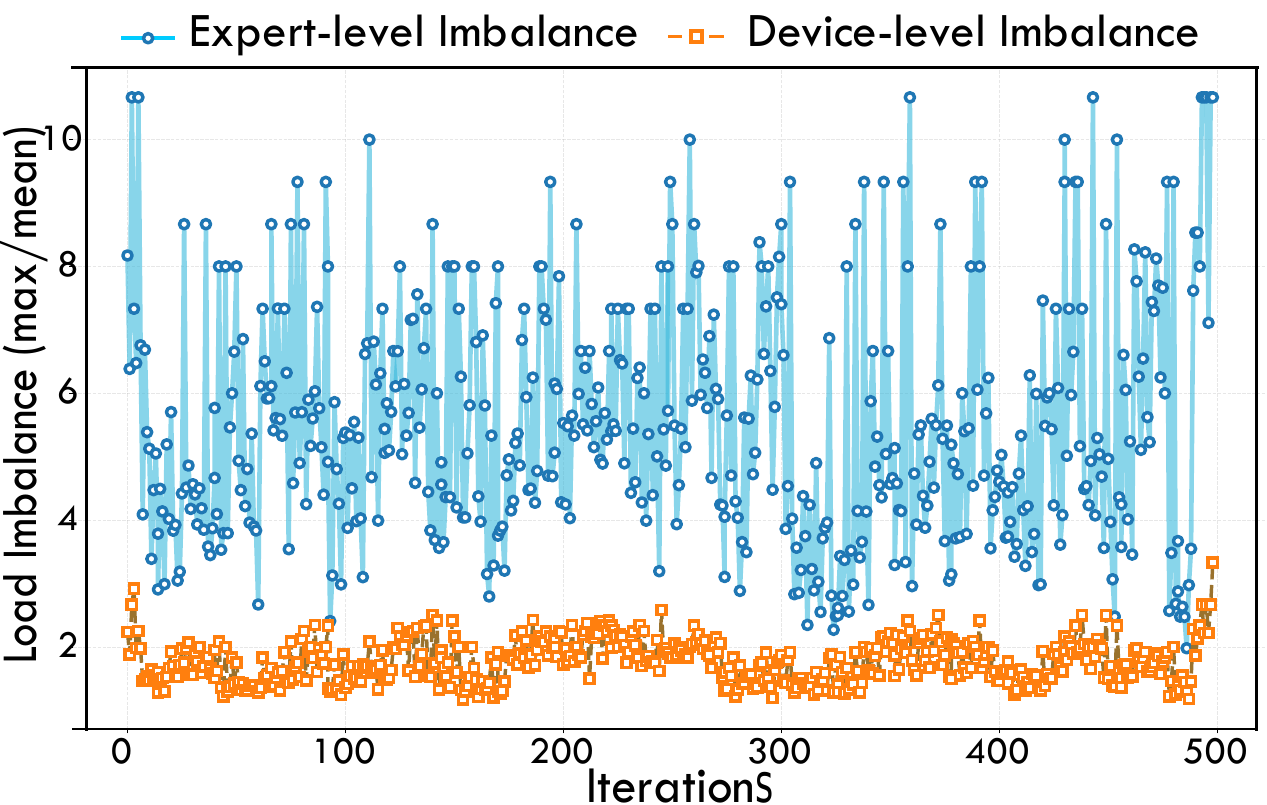}
        \caption{}
        \label{fig:maxload-iter}
    \end{subfigure}
    \hfill
    \begin{subfigure}{0.33\linewidth}
        \centering
        \includegraphics[width=\linewidth]{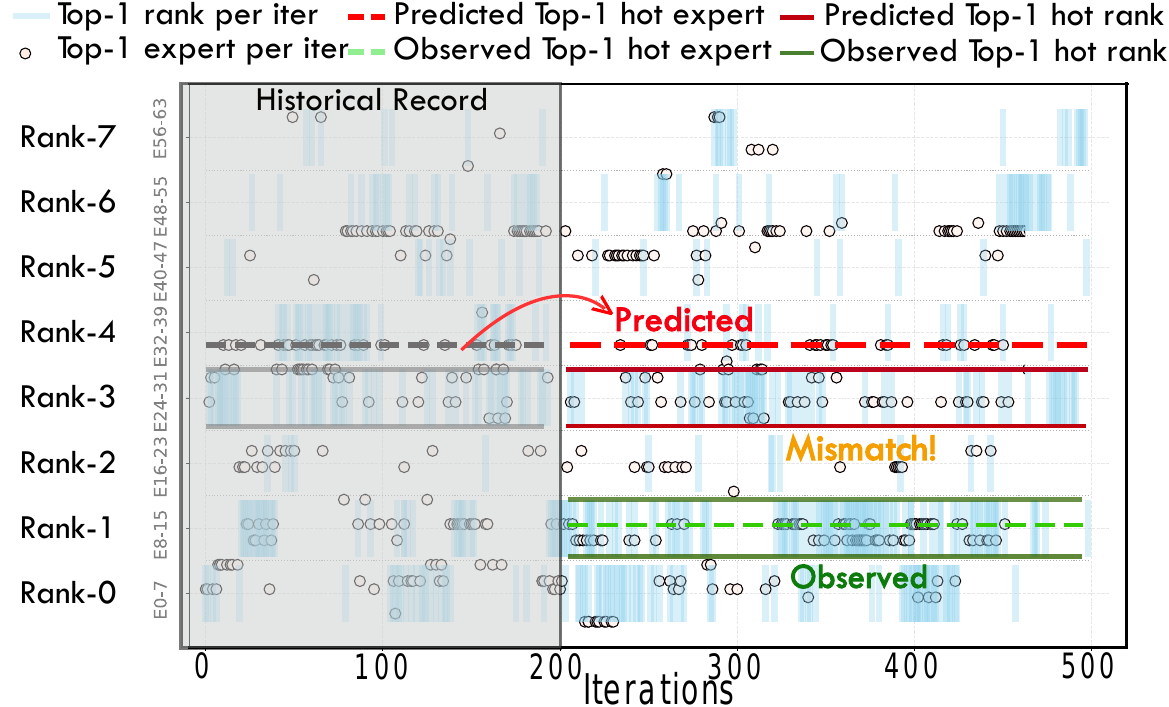}
        \caption{}
        \label{fig:maxload-dev-iter}
    \end{subfigure}
    \caption{High dynamics in MMoE routing limit the effectiveness of prediction-based methods. (a) Load imbalance across devices, experts, and modalities. (b) Temporal variation in imbalance severity across iterations. (c) Rapid shifts of the top-1 hot device and expert, highlighting the mismatch between historical observations and current load.}
\end{figure}

\paragraph{Routing Dynamics across Devices, experts, and Modalities.}
\Cref{fig:device-dynamics} shows that several ranks (e.g., $Rank_{0}$, $Rank_{1}$, and $Rank_{2}$) process more tokens than others.
Under EP execution, such load skew directly leads to stragglers, since MoE computation is globally synchronized across ranks.
At the expert level, the imbalance is further amplified: hot experts such as $E_{4}$ and $E_{5}$ on $Rank_{0}$ handle $5\sim6\times$ more tokens than the average expert.

This imbalance is largely driven by modality composition of the input.
In typical vision language tasks, high-resolution images produce a large number of vision tokens, while text prompts remain relatively short.
However, the proportion of vision tokens varies significantly across devices.
For instance, more than 90\% of tokens on $Rank_{0}$ are vision tokens, whereas on $Rank_{1}$ and $Rank_{4}$, this ratio stays below 50\%.
The heterogeneity is even more pronounced at the expert level.
Across different ranks, the top-1 hot expert exhibits widely varying modality composition, with the fraction of vision tokens ranging from 31\% to 93\%.
This indicates that modality-induced skew exhibits strong device-level dynamics, driven by highly variable routing behaviors at finer granularity.
Overall, this highlights two key observations:
\textit{\ding{172} vision tokens dominate the overall workload during multimodal inference, and their distribution is highly uneven across devices;
\ding{173} stragglers are primarily determined by which ranks aggregate vision-heavy experts.}


\paragraph{Routing Dynamics across Iterations.}

\Cref{fig:maxload-iter} shows that load imbalance fluctuates rapidly during inference. Within a short window, the ratio between the most-loaded expert and the average expert ranges from approximately 2$\times$ to 12$\times$. At the device level, the peak load often exceeds the average by more than 2$\times$, and in some cases, by over 3$\times$.
Notably, this device-level imbalance becomes more pronounced as the EP scale increases, since distributing experts across more devices increases routing skewness and amplifies the straggler effect.
Unlike MoE training, which exhibits strong temporal locality~\cite{fastermoe}, inference routing varies significantly across iterations. 
This behavior makes future load distributions difficult to predict, thereby fundamentally limiting the effectiveness of static or prediction-based load-balancing strategies.

\subsection{Limitations of Prediction-based Load Balancing}

We further analyze the limitations of prediction-based load balancing for MoE inference, using EPLB~\cite{eplb} as a representative example.

\paragraph{Prediction Mismatch.}
EPLB uses a sliding window over past iterations to maintain a global view of expert loads and periodically rebalances experts across EP ranks. For example, with a window size of 200 and a rebalancing interval of 300 iterations, hot devices are identified from the average load in recent iterations, and overloaded experts are replicated accordingly.
However, as shown in \Cref{fig:maxload-dev-iter}, such history-based predictions fail to track the highly dynamic load in MMoE inference. While $E_{33}$ and $Rank_{3}$ are top-1 hot spots during the first 200 iterations, in the subsequent 300 iterations, $Rank_{1}$ and $E_{11}$ become the hottest. 
These rapid fluctuations invalidate historical statistics, leading to delayed or ineffective rebalancing that not only fails to eliminate stragglers, but can even \textit{worsen} load imbalance when outdated predictions trigger incorrect replication decisions.

\paragraph{System Overhead.}
Prediction-based load balancing introduces both communication and memory overhead. 
For EPLB-like methods, each rebalancing step requires redistributing experts across devices, incurring communication cost. 
Let $K$ denote the number of expert replicas whose assignments differ between $ExpertMap_{\text{before}}$ and $ExpertMap_{\text{after}}$. 
The communication cost is $K \cdot Bytes_{\text{expert}}$, scaling linearly with the number of relocated replicas.
In large-scale EP deployments, this cost becomes substantial, especially when rebalancing is triggered frequently to track rapidly changing inference workloads, introducing non-trivial latency that can offset the expected benefits.

In addition, these methods increase memory usage due to redundant expert replicas. The per-rank memory overhead grows linearly with the number of redundant experts and MoE layers. 
For DeepSeek-V3, adding a single redundant expert per EP rank introduces approximately 2.4~GB of additional memory~\cite{vllm_eplb_params}, directly limiting the achievable batch size and overall inference throughput.




\subsection{Problem Formulation of Real-Time Load Balancing}
\label{sec:formu}
As observed above, MMoE inference exhibits highly dynamic routing, where expert loads and device imbalance can change rapidly across layers and iterations. 
Let $T_i(x)$ denote the execution time of EP rank $i$ for an MoE layer under routing outcome $x$. 
The layer latency is determined by the straggler:
\[
T_{\text{MoE}}(x) = \max_i T_i(x).
\]

The goal of load balancing is to reduce the straggler latency while accounting for its own overhead. 
We formulate the load balancing objective as:
\[
\max_{\pi} \;\; \mathbb{E}_x \big[ \max_i T_i(x) - \max_i T_i(\pi(x)) - T_{\text{LB}}(\pi, x) \big],
\]
where $\pi$ is the load balancing policy applied online based on the current routing $x$, and $T_{\text{LB}}$ denotes the latency overhead of applying $\pi$. 

Prediction-based methods approximate $\pi$ using historical statistics, implicitly assuming temporal locality. 
However, rapid routing dynamics violate this assumption, making prediction unreliable even with frequent updates while incurring extra communication cost. 
Moreover, optimizing proxy metrics such as token or expert balance is not a necessary condition for minimizing $\max_i T_i$. 
Although such proxies may correlate with execution latency, enforcing them incurs significant overhead that can offset the gains. 
Furthermore, unlike training where token balance supports optimization stability, inference focuses solely on minimizing execution latency, making token balance an indirect and non-essential objective.

Therefore, an effective policy $\pi$ must be both real-time and lightweight: it should (i) operate online based on the \textit{current} routing outcome, (ii) directly target device-level \textit{latency imbalance}, and (iii) introduce negligible overhead so that the \textit{net} latency reduction remains positive.
\section{Methodology}
\label{sec:method}

\subsection{System Overview}
\label{sec:overview}

\begin{figure}[th]
    \centering
    \includegraphics[width=0.96\linewidth]{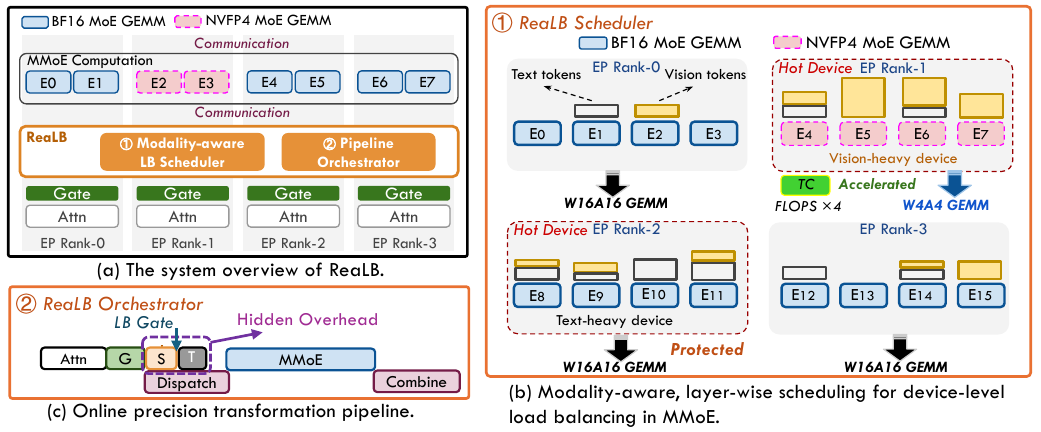}
    \caption{ReaLB design for efficient multimodal MoE inference. It combines modality-aware scheduling and an overlapping execution pipeline to mitigate device-level latency imbalance while minimizing runtime overhead.}
    \label{fig:overview}
\end{figure}

We propose ReaLB to reduce rank-wise latency imbalance in MMoE inference while minimizing runtime scheduling overhead. 
ReaLB is integrated into existing multimodal LLM inference frameworks such as vLLM~\cite{vllm} and SGLang~\cite{sglang}, and performs online load balancing at the granularity of EP ranks.
The overall workflow of ReaLB is illustrated in \Cref{fig:overview}(a), and the inference pipeline consists of four stages:

\begin{itemize}
\itemsep0em

\item[\ding{172}] \textbf{Routing and Profiling:} Input tokens are processed by attention and gating modules, producing real-time expert assignments and device-level load statistics.

\item[\ding{173}] \textbf{Modality-aware LB Scheduling:} The scheduler analyzes current routing outcomes to detect device-level imbalance. It identifies overloaded ranks, often caused by skewed assignment of vision-heavy experts, and constructs a per-rank execution plan by assigning GEMM precision (\S\ref{subsec:lb}). This enables straggling ranks to accelerate execution using low-precision hardware units.

\item[\ding{174}] \textbf{Overhead-aware Pipeline Orchestration:} The orchestrator executes the scheduling plan by collecting runtime metadata and performing online precision transformation (e.g., BF16 to FP4). These operations are overlapped with inter-device communication (e.g., \texttt{All-to-All Dispatch}), hiding transformation overhead from the critical path (\S\ref{subsec:overlap}).

\item[\ding{175}] \textbf{Balanced MoE Execution:} Each EP rank executes MMoE computation according to the generated plan, leveraging FP4 Tensor Cores for  precision-insensitive workloads, thereby reducing execution skew across ranks.

\end{itemize}

\subsection{Real-time Modality-aware LB Scheduling}
\label{subsec:lb}

We model ReaLB as a feedback control policy that adjusts execution precision based on instantaneous routing states in MMoE inference.
The policy operates online before each MoE layer without relying on historical statistics, enabling adaptation to rapidly changing routing patterns.

\paragraph{Runtime State.}
We define the device-level imbalance as
\[
IB_d = \frac{Load_d}{Ideal}, \quad IB_{\text{global}} = \max_d IB_d,
\]
where $Load_d$ denotes the token load on device $d$, and $Ideal$ is the average load across EP ranks at current step.
Each device is further characterized by the vision token ratio
\[
R_{vd} = \frac{N_{vd}}{N_{vd}+N_{td}},
\]
where $N_{vd}$ and $N_{td}$ denote the numbers of vision and text tokens on device $d$.
The runtime state for each device is thus defined by $(IB_{\text{d}}, R_{vd})$.

\paragraph{Hierarchical Control Policy.}
ReaLB adopts a two-stage control policy applied synchronously at each MMoE layer.
First, devices exhibiting significant imbalance are identified as hotspots according to:
\[
\mathcal{H} = \{ d \mid IB_d > C \},
\]
where $C=1$ corresponds to a perfectly balanced state. This step defines the set of overloaded devices within the current layer.
Given the hotspot set $\mathcal{H}$, ReaLB performs device-level precision assignment based on modality skew. Devices satisfying
\[
R_{vd} > M_d, \quad d \in \mathcal{H},
\]
are marked as vision-heavy and compressible, and their hosted experts are executed using low-precision GEMM kernels, while all remaining devices execute in standard precision.

\paragraph{Adaptive Control via AIMD.}
We adopt an additive-increase multiplicative-decrease (AIMD) control principle~\cite{aimd}, a classical feedback mechanism widely used in congestion control systems.
AIMD provides a stable trade-off between responsiveness under high imbalance and stability under normal conditions.
The control parameter $M_d$ is updated based on the global imbalance signal:
\[
M_d =
\begin{cases}
0.5 M_d, & \text{if } IB_{\text{global}} > \tau, \\
\min(1, M_d + 0.1), & \text{otherwise},
\end{cases}
\]
where $\tau$ denotes the congestion threshold, which is set to $1.5$ by default.
This rule increases sensitivity to modality skew under severe congestion, while gradually relaxing constraints when the system stabilizes.
A detailed empirical analysis of the evolution of $M_d$ is provided in Appendix~\ref{app:aimd}.

\paragraph{Scheduling Effect.}
As shown in \Cref{fig:overview}(b), both $Rank_1$ and $Rank_2$ are identified as overloaded devices, while only $Rank_1$ exhibits strong vision dominance.
At runtime, ReaLB applies FP4 GEMM execution to $Rank_1$ according to the control policy, while preserving standard precision for other ranks.
This selective execution targets routing-induced hotspots, which are the primary contributor to MMoE latency imbalance.

\subsection{Overhead-Aware Pipeline Orchestration}
\label{subsec:overlap}

To minimize LB overhead, ReaLB introduces an orchestrator that operates through two key mechanisms: (i) fine-grained overlap between LB scheduling and communication, and (ii) a lightweight LB gate that controls runtime activation, as illustrated in \Cref{fig:overview}(c).

\paragraph{Overlapping Pipeline.}
To support real-time precision switching without increasing memory footprint, ReaLB performs \emph{on-the-fly precision transformation}.
Each expert stores only its original high-precision weights alongside precomputed scaling factors required for low-precision conversion.
When the scheduler assigns a new precision configuration, the selected expert groups are quantized online to the target precision and immediately launched for computation.
This design eliminates the need for maintaining multiple precision copies of expert weights.

While online transformation introduces additional computation, ReaLB is designed to mask this overhead entirely.
ReaLB manages two primary sources of runtime latency:
\ding{172} \textit{Scheduling Metadata (S):} Routing statistics are collected across EP ranks via a lightweight \texttt{allgather}.
\ding{173} \textit{Precision Transformation (T):} The transformation latency of weights from high to low precision (e.g., BF16 to FP4).
Under the standard DP attention and EP MoE deployment, tokens are dispatched to their target devices after gating, and MoE computation is then executed sequentially.
Instead, ReaLB adopts a pipelined execution design that overlaps metadata collecting and weights quantization with inter-device communication (e.g., \texttt{dispatch}), which dominates latency in practice.

As a result, the LB cost does not scale with the EP size, since it only involves lightweight metadata collection and local precision transformation. In contrast, \texttt{dispatch} latency increases with EP scale and becomes further amplified under multi-node deployment. This ensures that LB overhead is fully hidden by communication, making the overall system scalable.

\begin{wrapfigure}{r}{0.38\linewidth}
    \centering
    \includegraphics[width=\linewidth]{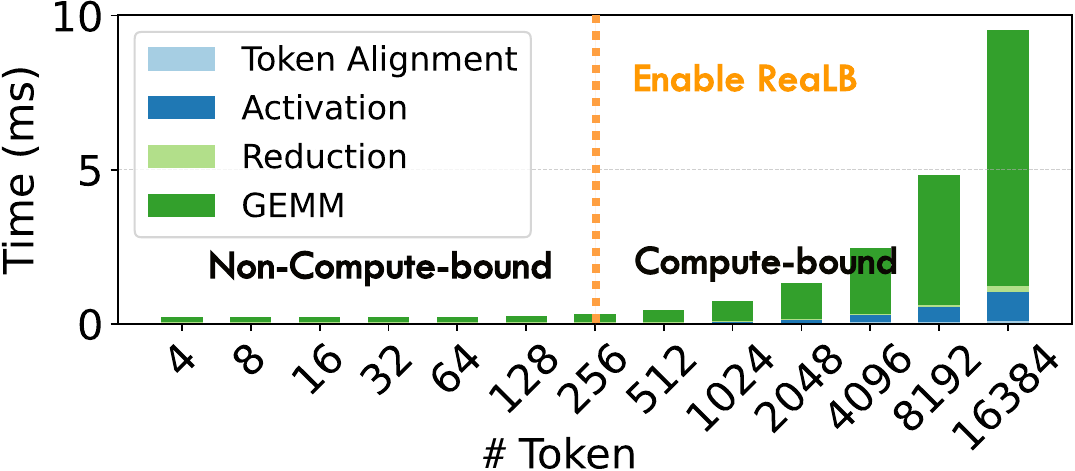}
    \caption{ReaLB is enabled only in the compute-bound regime
with large batch sizes.}
    \label{fig:batch_threshold}
\end{wrapfigure}

\paragraph{LB Gate.}
To precisely control runtime overhead, we introduce an LB gate that operates after routing metadata collection to determine whether load balancing is activated.
ReaLB targets the regime where MoE execution is dominated by GEMM computation, such that device-level load imbalance directly translates into latency imbalance. The activation condition is based on a global batch threshold $\Gamma$, where ReaLB is enabled when the aggregated load across devices exceeds $\Gamma$.
As shown in \Cref{fig:batch_threshold}, when per-rank token counts are small, non-GEMM operations dominate latency and device imbalance has limited impact; as batch size increases, GEMM becomes the dominant cost and straggler effects emerge as the primary bottleneck.
Only under this regime is the LB gate enabled; otherwise, ReaLB falls back to standard execution, ensuring negligible $T_{\text{LB}}$ in the overall latency objective.

\section{Evaluation}
\label{eval}

\subsection{Setup}
\label{sec:setup}

\textbf{Implementation Details.}
We build ReaLB on vLLM v0.13.0~\cite{vllm}, with mixed-precision MMoE support enabled via LLM Compressor~\cite{llmcompressor} and NVFP4 GEMM kernels from FlashInfer~\cite{flashinfer}. The system is developed in PyTorch and Triton.
We adopt a prefill–decode (PD) colocated deployment, and use vLLM’s default continuous batching scheduler for serving. Communication and computation are overlapped using CUDA streams. All experiments are conducted on eight NVIDIA RTX 5090 GPUs under combined data and expert parallelism.
With the colocated deployment and continuous batching mechanism in vLLM, each forward batch may contain a mixture of prefill and decode requests. 
Additional routing statistics and batching patterns are provided in Appendix~\ref{sec:append_routing} and ~\ref{sec:append_batching}.
We set the capacity factor $C=1$, the modality threshold $M_d$ to 0.9 initially, and the global batch threshold $\Gamma=2048$ ($256 \times 8\ ranks$).

\textbf{Reproducibility.}
To facilitate reproducibility, we provide the implementation of ReaLB and the associated scripts in the supplementary material.

\textbf{Models.}  
We evaluate two recent open-source multimodal MoE models: Kimi-VL~\cite{kimi} and Qwen3-VL~\cite{qwen3-vl} with 64 and 128 routed experts, respectively. 
We further apply NVFP4 post-training quantization (PTQ)~\cite{llmcompressor} to the MoE layers to obtain scale factors for mixed-precision execution.

\textbf{Compared Methods.}
To study performance and balance trade-offs, we compare the following configurations:
    (1) \textbf{Baseline:} Standard W16A16 GEMM execution for all MoE layers without load balancing.
    (2) \textbf{FP4-All:} W4A4 NVFP4 GEMM execution for all MoE layers.
    (3) \textbf{EPLB:} A history‑based expert‑placement LB method (following \cite{eplb}).
    (4) \textbf{Async\_EPLB:}  An asynchronous version of EPLB that overlaps weight transfers to reduce replacement overhead.
    (5) \textbf{ReaLB:} Our proposed modality‑aware real‑time LB scheme with overlapped pipeline, referred to as the $full$ ReaLB.
    (6) \textbf{ReaLB-seq:} A sequential variant of ReaLB that disables pipeline overlapping.
    (7) \textbf{ReaLB-m$i$:} ReaLB with fixed modality threshold $M_d$ that disable \textit{AIMD-based} adaptive control.

\textbf{Metrics.}
We evaluate model accuracy on representative multimodal benchmarks using \texttt{lmms-eval}~\cite{lmms}. 
Efficiency is measured on the same workloads to ensure consistency between accuracy and system evaluation. 
We use CUDA events for fine-grained timing and the vLLM benchmark for end-to-end throughput, measured in processed tokens per second (input + output).
To study the overall performance, we report three metrics: (i) accuracy degradation ($\Delta$Acc) relative to the BF16 baseline, (ii) end-to-end throughput speedup relative to the BF16 baseline, and (iii) MoE layer latency to capture per-layer execution efficiency.

\subsection{Main Results and Trade-offs}
\label{eval:main}
We compare ReaLB against representative baselines, including EPLB as a prediction-based non-real-time LB method and FP4-All as a uniform low-precision quantization scheme. We further evaluate two fixed modality-threshold variants, ReaLB-m1 ($M_d=0$) and ReaLB-m2 ($M_d=0.7$), on the two MMoE models using MMMU~\cite{mmmu}, MathVista~\cite{MathVista}, and DynaMath~\cite{DynaMath}.
End-to-end speedup and accuracy degradation ($\Delta$Acc) are reported relative to the BF16 baseline.
As shown in \Cref{tab:tradeoff_analysis}, ReaLB consistently improves end-to-end throughput across all benchmarks while maintaining a favorable accuracy–efficiency trade-off compared to existing methods.

\definecolor{good}{RGB}{230,240,255}
\definecolor{highlight}{RGB}{224, 224, 224} 

\begin{table*}[thb]
\centering
\small
\setlength{\tabcolsep}{3pt}
\caption{Main performance comparison and trade-offs under different load balancing strategies. The blue-highlighted values indicate the best results among ReaLB variants (gray-shaded columns).}
\label{tab:tradeoff_analysis}

\begin{tabular}{c|c|ccccccc}
\toprule
\textbf{Model} & \textbf{Workload} 
& Baseline & EPLB & FP4-All 
& \cellcolor{highlight} ReaLB-m1 
& \cellcolor{highlight} ReaLB-m2 
& \cellcolor{highlight} ReaLB-seq 
& \cellcolor{highlight} ReaLB \\
\midrule

\multirow{6}{*}{\textbf{Kimi-VL}}
& MMMU $\Delta$Acc $\downarrow$     & 0.00 & 0.00 & -4.22 & -2.22 & \cellcolor{good} \textbf{0.00} & -1.22 & -1.22 \\
& Speedup  $\uparrow$             & 1.00 & 0.91 & 1.34 & 1.32 & 1.29 & 1.25 & \cellcolor{good} \textbf{1.32} \\
\cmidrule{2-9}

& MathVista $\Delta$Acc $\downarrow$ & 0.00 & 0.00 & -2.00 & -2.00 & \cellcolor{good} \textbf{-1.10} & -1.70 & -1.70 \\
& Speedup   $\uparrow$             & 1.00 & 0.95 & 1.40 & \cellcolor{good} \textbf{1.32} & 1.30 & 1.26 & 1.31 \\
\cmidrule{2-9}

& DynaMath $\Delta$Acc $\downarrow$  & 0.00 & 0.00 & -8.19 & -3.80 & \cellcolor{good} \textbf{-2.50} & -3.10 & -3.10 \\
& Speedup   $\uparrow$             & 1.00 & 0.96 & 1.41 & 1.30 & 1.29 & 1.25 & \cellcolor{good} \textbf{1.30} \\

\midrule\midrule

\multirow{6}{*}{\textbf{Qwen-VL}}
& MMMU $\Delta$Acc $\downarrow$    & 0.00 & 0.00 & -2.44 & -0.34 & -0.34 & -0.00 & \cellcolor{good} \textbf{-0.00} \\
& Speedup   $\uparrow$            & 1.00 & 0.91 & 1.18 & 1.13 & 1.13 & 1.01 & \cellcolor{good} \textbf{1.13} \\
\cmidrule{2-9}

& MathVista $\Delta$Acc $\downarrow$ & 0.00 & 0.00 & -2.70 & -1.60 & -1.30 & -0.40 & \cellcolor{good} \textbf{-0.40} \\
& Speedup  $\uparrow$              & 1.00 & 0.83 & 1.16 & \cellcolor{good} \textbf{1.14} & 1.13 & 1.01 & 1.13 \\
\cmidrule{2-9}
& DynaMath $\Delta$Acc $\downarrow$ & 0.00 & 0.00 & -1.89 & -1.59 & -1.09 & -0.39 & \cellcolor{good} \textbf{-0.39} \\
& Speedup   $\uparrow$         & 1.00 & 0.84 & 1.12 & 1.10 & 1.10 & 1.02 & \cellcolor{good} \textbf{1.10} \\

\bottomrule
\end{tabular}
\end{table*}

\begin{wraptable}{r}{0.55\linewidth}
\centering
\small
\setlength{\tabcolsep}{3pt}
\caption{Accuracy stability of ReaLB across additional multimodal benchmarks ($\Delta$Acc $\downarrow$).}
\label{tab:acc}
\vspace{-0.5em}

\begin{tabular}{c|cccc|c}
\toprule
Model &
AI2D & InfoVQA & TextVQA & MMBench & AVG \\
\midrule

Kimi-VL
& -0.13 & -0.38 & -0.91 & -0.00 & -0.36 \\

\midrule

Qwen-VL
& -1.23 & -0.00 & -0.53 & -0.60 & -0.59 \\

\bottomrule
\end{tabular}

\vspace{-0.8em}
\end{wraptable}

\textbf{Comparison with prior methods.}
We first compare ReaLB with prior approaches. EPLB fails to provide consistent benefits and can even degrade performance, leading to negative speedup in some settings. This is mainly due to the mismatch between historical token statistics and dynamic routing behavior in multimodal MoE, which limits its ability to capture real-time hotspots and introduces additional communication and expert replacement overhead.
Compared to uniform low-precision execution (FP4-All), ReaLB achieves a significantly better trade-off. On Kimi-VL, it achieves up to 1.32$\times$ speedup on MMMU and 1.30$\times$ on DynaMath, while substantially reducing accuracy degradation. In contrast, FP4-All introduces notable accuracy loss (e.g., -8.19\% on DynaMath), indicating that uniform quantization is overly aggressive for multimodal workloads. ReaLB reduces accuracy loss to 3.1\% on DynaMath while maintaining comparable throughput gains. 
\Cref{tab:acc} further validates the stability of ReaLB in accuracy preservation across additional multimodal benchmarks.
This demonstrates that exploiting modality heterogeneity and precision sensitivity differences under runtime-aware control leads to a better trade-off than uniform quantization.

\textbf{Robustness of modality-aware design across models.}
We observe that the benefit of ReaLB generalizes across different multimodal MoE backbones. On Qwen-VL, ReaLB achieves near-lossless accuracy (with <0.6\% average degradation) while delivering 1.10$\times$–1.13$\times$ speedup across benchmarks. Together with Kimi-VL results, this confirms that modeling modality heterogeneity and runtime imbalance enables stable and consistent improvements across architectures.

\textbf{Ablation and sensitivity analysis.}
To analyze the contribution of each component in ReaLB, we conduct ablation studies. ReaLB-seq removes pipeline overlap while retaining modality-aware scheduling, resulting in similar accuracy but reduced speedup, which highlights the importance of overlapping scheduling and precision transformation with communication. In addition, sensitivity analysis on $M_d$ shows that adaptive control provides more stable performance than static thresholds, achieving a better balance between aggressiveness and accuracy preservation.

\begin{figure}[t]
    \centering

    \begin{subfigure}{0.35\linewidth}
        \centering
        \includegraphics[width=\linewidth]{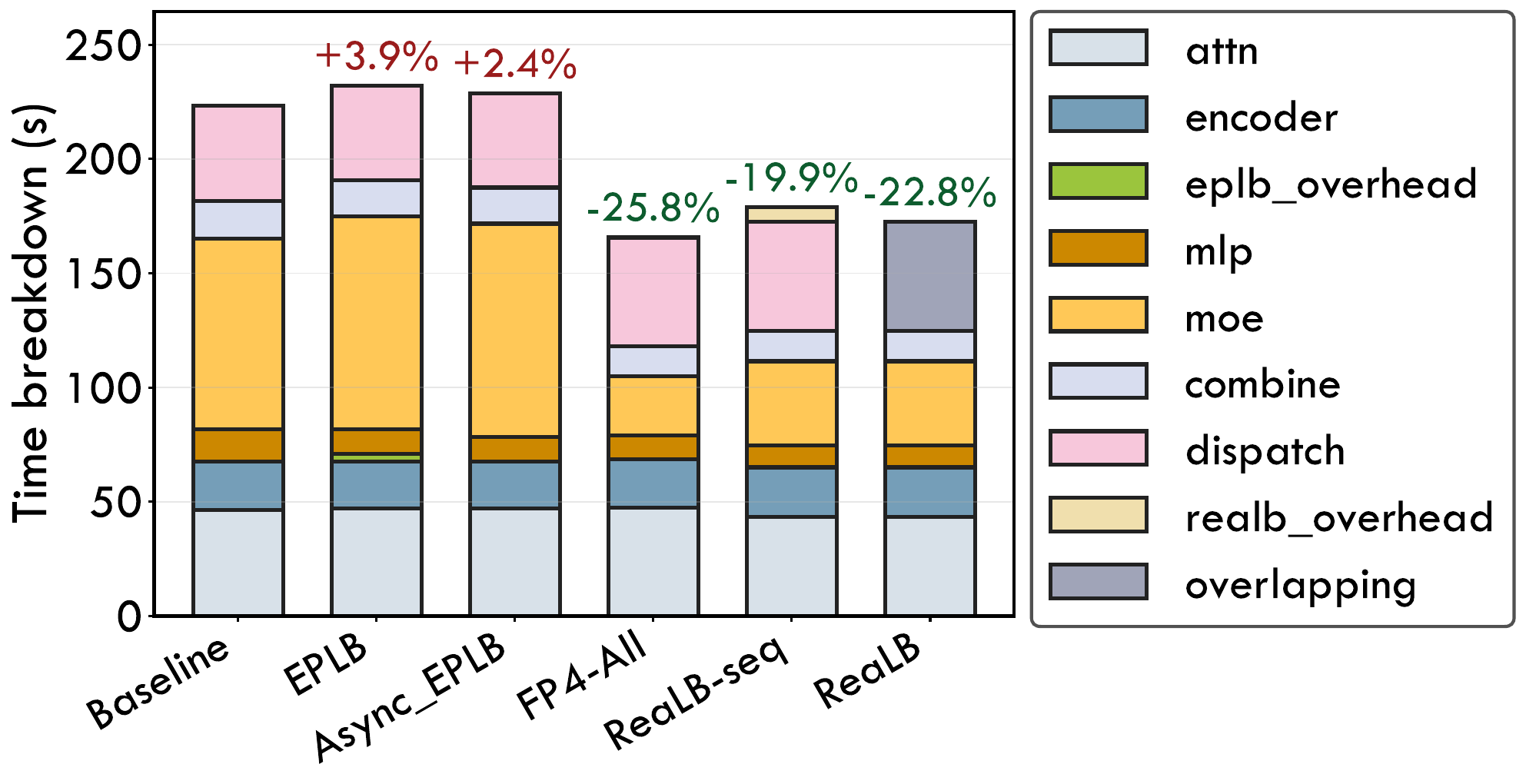}
        \caption{End-to-end time breakdown.}
    \end{subfigure}
    \hfill
    \begin{subfigure}{0.3\linewidth}
        \centering
        \includegraphics[width=\linewidth]{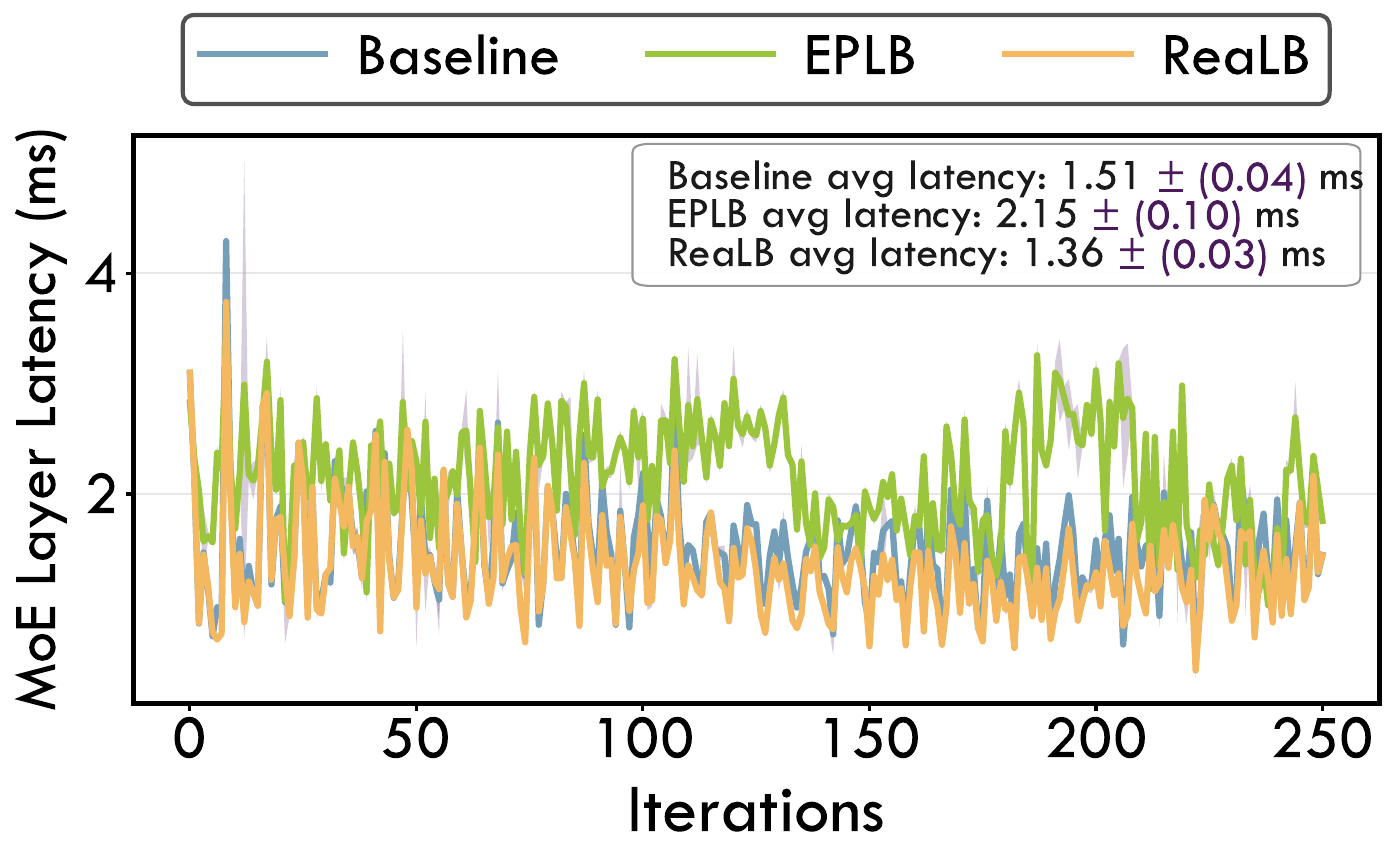}
        \caption{MoE latency over iterations.}
    \end{subfigure}
    \hfill
    \begin{subfigure}{0.32\linewidth}
        \centering
        \includegraphics[width=\linewidth]{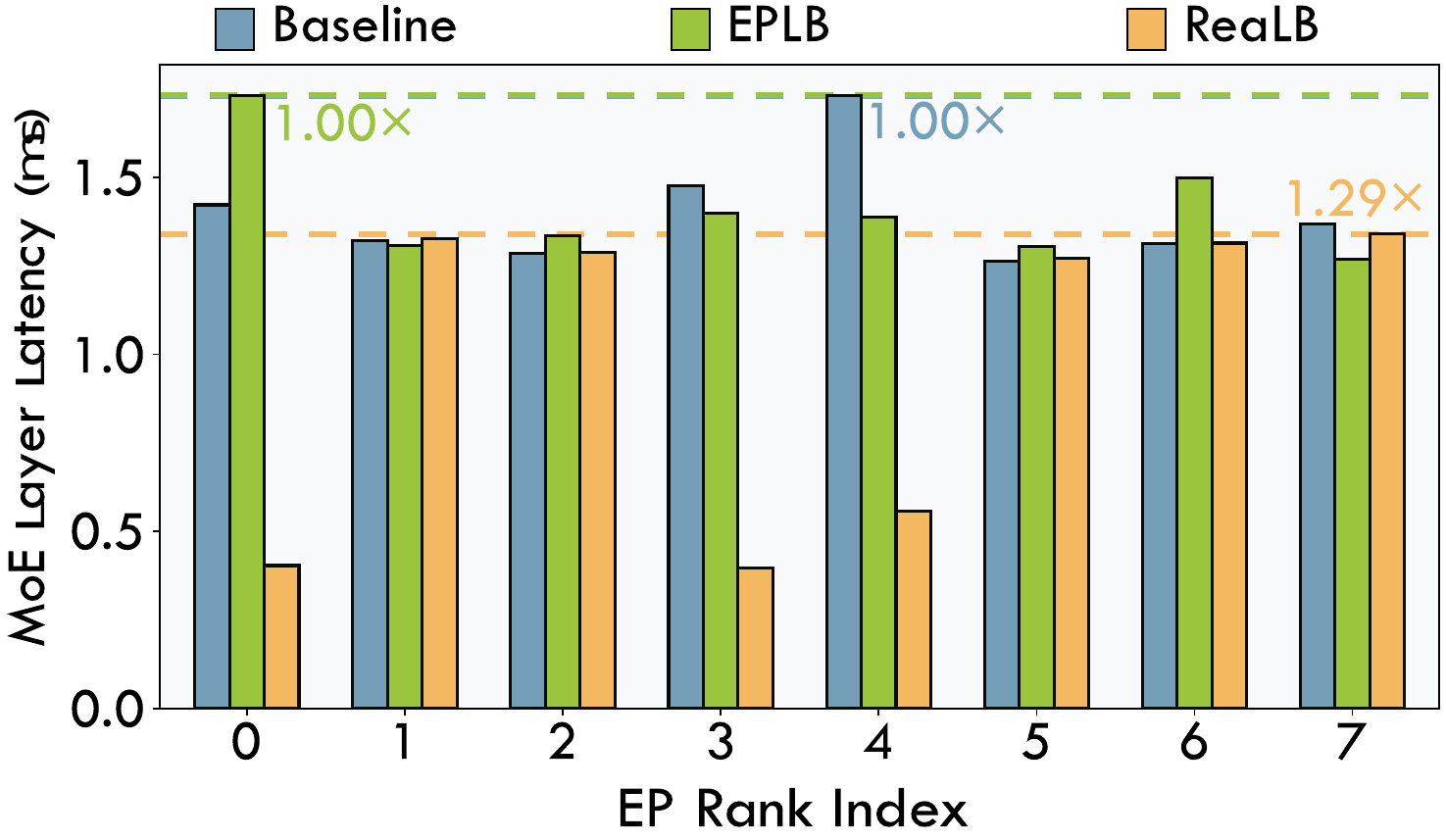}
        \caption{Rank-wise latency distribution.}
    \end{subfigure}

    \vspace{0.5em}

    \begin{subfigure}{0.35\linewidth}
        \centering
        \includegraphics[width=\linewidth]{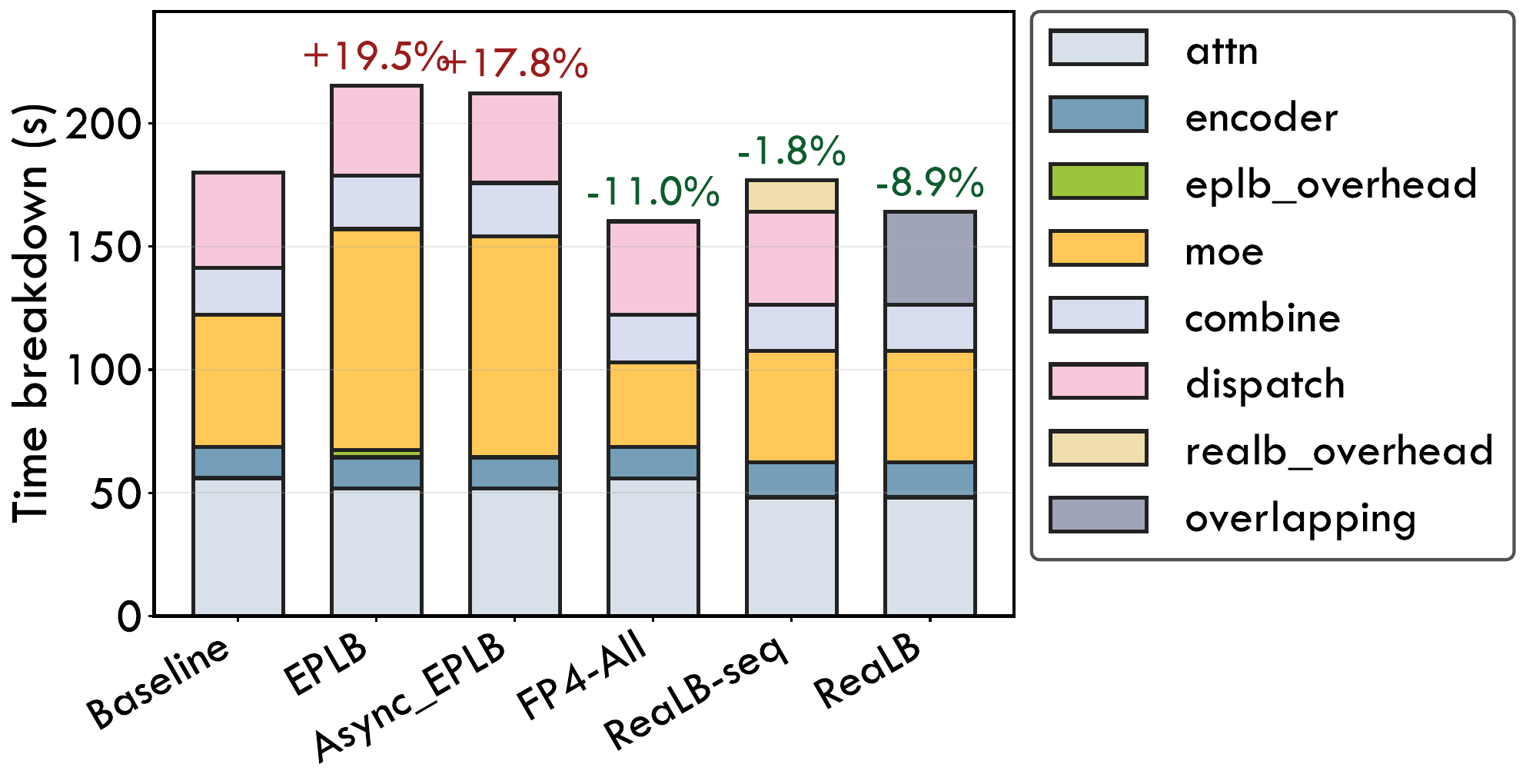}
        \caption{End-to-end time breakdown.}
    \end{subfigure}
    \hfill
    \begin{subfigure}{0.3\linewidth}
        \centering
        \includegraphics[width=\linewidth]{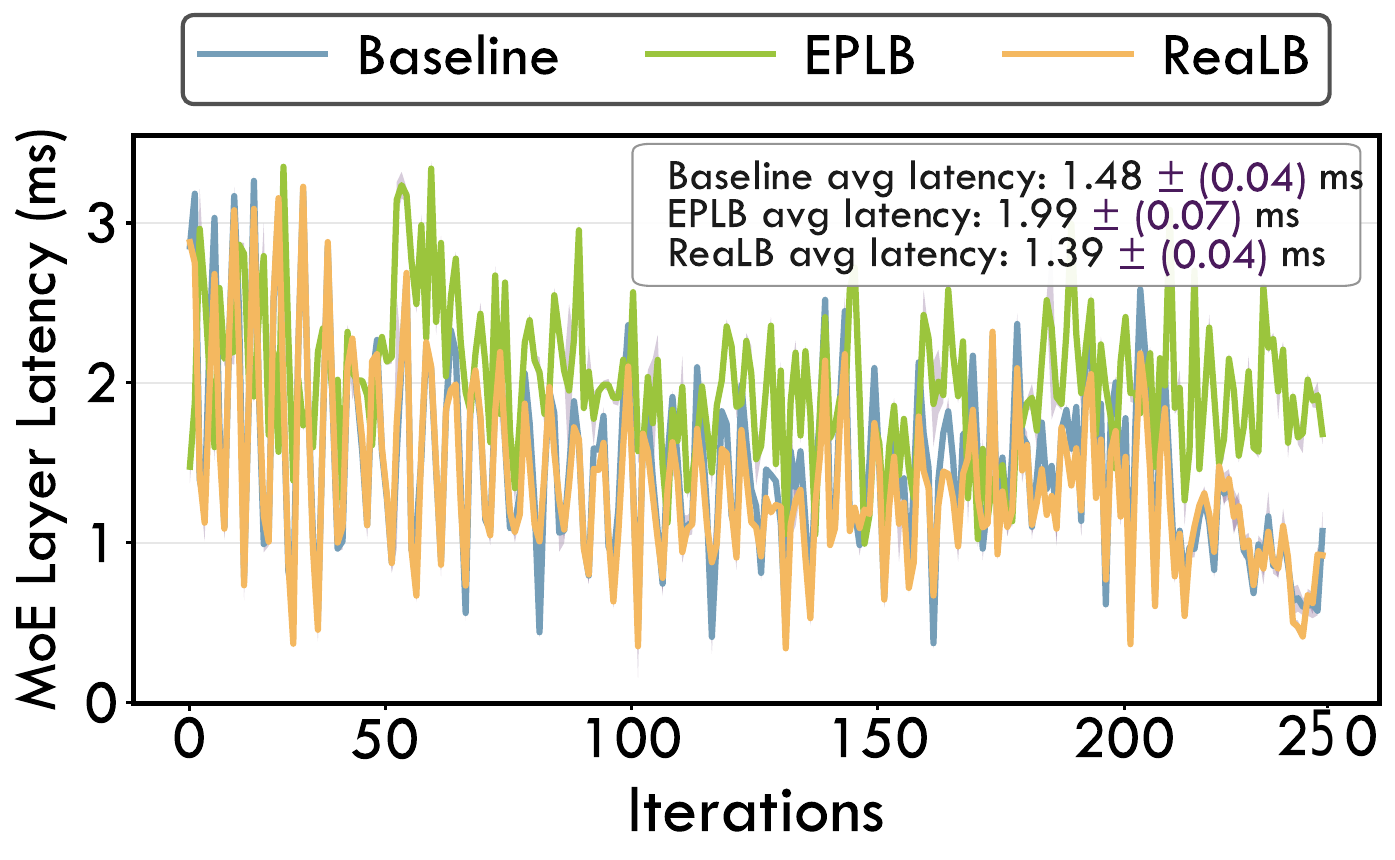}
        \caption{MoE latency over iterations.}
    \end{subfigure}
    \hfill
    \begin{subfigure}{0.32\linewidth}
        \centering
        \includegraphics[width=\linewidth]{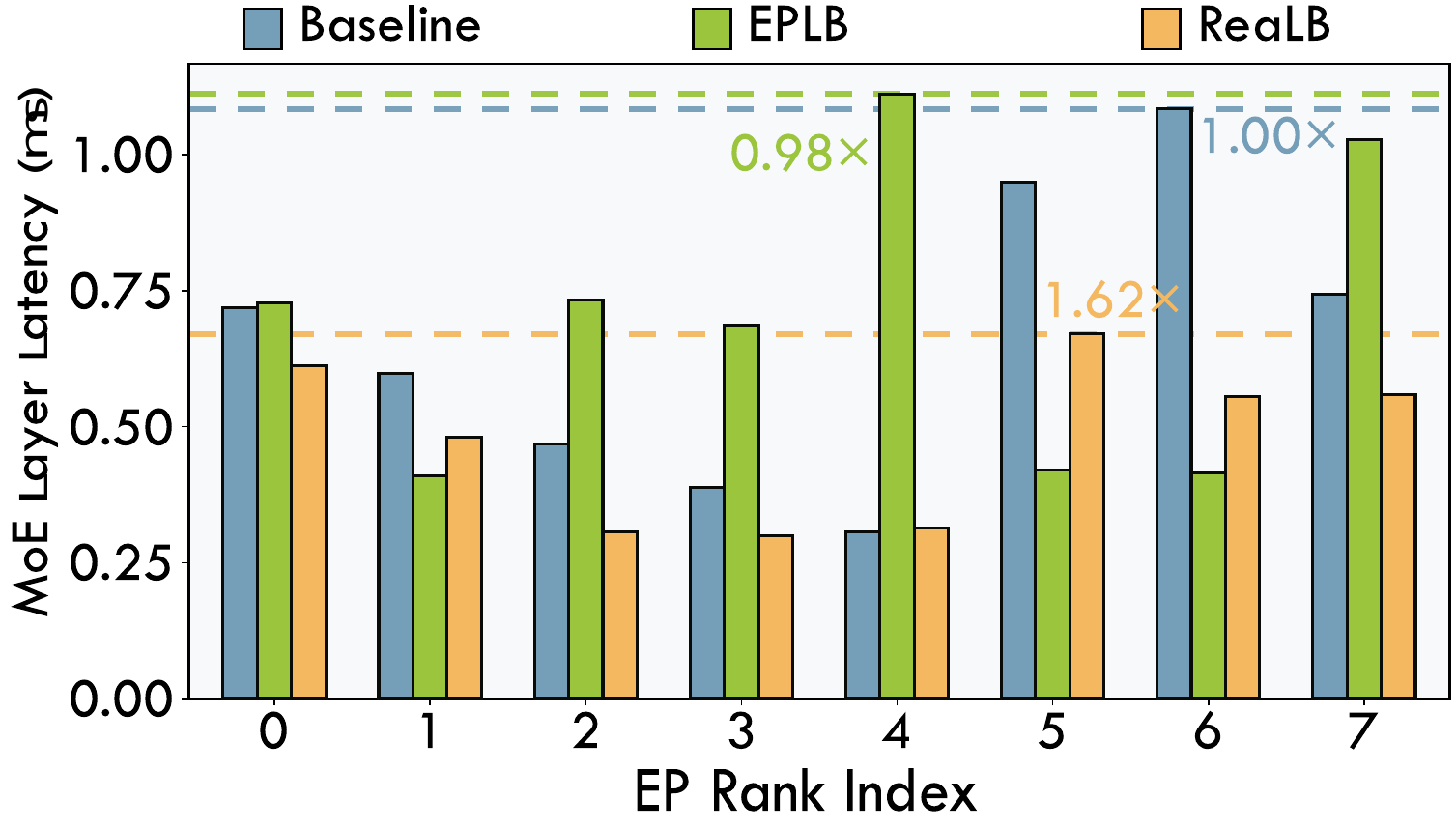}
        \caption{Rank-wise latency distribution.}
    \end{subfigure}

    \caption{Fine-grained MoE latency analysis. Top row (a–c): Kimi-VL; bottom row (d–f): Qwen-VL.}
    \label{fig:module_breakdown}
\end{figure}

\subsection{Latency Breakdown.}
We further conduct fine-grained latency analysis on DynaMath (1K samples) under different load balancing strategies.
On Kimi-VL, \Cref{fig:module_breakdown}(a) shows that FP4-All and ReaLB nearly halve the MoE time share, yielding 22.8\% end-to-end time reduction, while EPLB slightly increases it and leads to negative gains. 
FP4-All serves as an upper bound, being 6\% faster than ReaLB-seq; meanwhile, ReaLB achieves 3\% higher speedup than ReaLB-seq, indicating that overhead control effectively removes pipeline bubbles and recovers the performance gap.
\Cref{fig:module_breakdown}(b,c) explain the source of these gains. 
ReaLB reduces average MoE latency from 1.51 ms to 1.36 ms, whereas EPLB increases it from 1.51 ms to 2.15 ms, showing the limitation of prediction-based schemes under dynamic workloads over the observed iterations. 
At the rank level, ReaLB achieves up to 1.29\(\times\) speedup by targeting hotspot ranks, confirming the effectiveness of runtime-aware load balancing.
From \Cref{fig:module_breakdown}(d–f), we observe a similar trend on Qwen-VL, where ReaLB consistently outperforms EPLB with a clear advantage in both latency reduction and stability.
Additional results are provided in Appendix \ref{app:efficiency_more}.

\section{Conclusion}
\label{sec:conclu}

We propose \textbf{ReaLB}, a real-time, modality-aware load balancing approach for multimodal MoE inference. 
Unlike prior prediction-based methods, ReaLB directly targets execution-time imbalance at the MoE layer, enabling accurate and timely load balancing without relying on indirect workload estimation. 
By selectively applying low-precision execution based on modality, ReaLB mitigates stragglers caused by dynamic and skewed routing patterns. 
An overlapped execution design further hides scheduling and precision conversion overhead, ensuring low runtime cost.
Experiments across multiple multimodal MoE models show that ReaLB reduces MoE latency while preserving accuracy, demonstrating the practicality of real-time load balancing for multimodal inference.

\bibliographystyle{plain}
\bibliography{ref}


\newpage
\appendix
\onecolumn
\onecolumn
\section{Broader Impacts}
\label{app:impacts}
\textbf{Positive societal impact.}
ReaLB improves multimodal MoE inference efficiency by reducing execution imbalance and improving hardware utilization under mixed-precision execution. This leads to higher device utilization and lower end-to-end latency, improving throughput efficiency in large-scale multimodal serving systems. By better aligning computation with runtime workload imbalance, ReaLB can also reduce redundant computation and improve energy efficiency during inference, which is important for cost-sensitive datacenter deployment of large multimodal models. In addition, improved load balancing enhances system stability under high concurrency, which is critical for production-level multimodal services.

\textbf{Negative societal impact.}
ReaLB may introduce potential risks related to accuracy degradation under mixed-precision execution. Although the observed accuracy loss is generally small in our evaluation, aggressive precision control or misconfiguration may lead to non-negligible performance degradation on certain inputs or tasks, particularly under distribution shifts. Such effects may be more critical in high-stakes applications, such as medical image analysis or other safety-critical multimodal systems, where even small accuracy variations can impact downstream decisions. Therefore, careful validation and conservative deployment are required when integrating ReaLB into such scenarios.

\section{LLM Usage Declaration}
\label{app:llm}
This manuscript uses LLMs strictly for the purpose of language editing and textual polishing to enhance presentation quality. We declare that the novel ideas, methodological framework, experimental execution, and data analysis are the original work of the authors. All content modified by AI tools has been carefully reviewed and validated by the authors to ensure accuracy.

\section{Compute Reporting}
\label{app:compute}
Reproducing the full experimental results reported in this paper requires approximately 5 days on 8 NVIDIA RTX 5090 GPUs.
This estimate includes all evaluation workloads and profiling experiments used in the paper.

\section{Limitations}
\label{app:limit}
ReaLB directly targets execution-time imbalance without explicit token redistribution, avoiding the mismatch introduced by prediction-based methods.
The effectiveness of ReaLB is closely related to the intrinsic routing characteristics of the underlying MoE model and the deployment configuration.
For models trained with strong load-balancing regularization (e.g., higher auxiliary loss weights), routing skew during inference can be naturally mitigated, which reduces the degree of execution imbalance that ReaLB can further optimize.
The degree of device-level imbalance is also influenced by expert parallelism (EP) configuration.
When the number of devices is small relative to the total number of experts (e.g., 2--4 GPUs with >64 routed experts), each device hosts a larger set of experts, which statistically smooths token distribution and alleviates inter-device imbalance.
In such cases, the optimization space for execution-time rebalancing becomes more limited.
In addition, ReaLB currently leverages mixed-precision execution for latency control. Extending this mechanism to richer precision configurations, together with corresponding kernel improvement, remains an important direction for future work.

\section{Implementation Details}
\label{app:impl}
\textbf{Models.} 
Our primary target is large-scale multimodal MoE models, such as Qwen3-VL-235B-A22B~\cite{qwen3-large}.
However, due to GPU memory constraints (8 GPUs with 32\,GB each), we evaluate ReaLB on relatively lightweight yet representative multimodal MoE models:
Kimi-VL-A3B-Instruct~\cite{kimivl_model}, 
Qwen3-VL-30B-A3B-Instruct~\cite{qwen3-vl}.
These models retain key characteristics of large-scale multimodal MoE systems, including sparse expert routing and mixed-modality token distributions.

\textbf{Datasets.}  
We evaluate multimodal MoE models on the following vision-language benchmarks using \texttt{lmms-eval}~\cite{lmms}.  
(1) \textbf{MMMU}~\cite{mmmu}: multi-image reasoning, requiring cross-image integration; 
The selected benchmarks cover diverse modality compositions:  
(2) \textbf{MathVista}~\cite{MathVista}: visual mathematical reasoning benchmark requiring integration of visual context and symbolic math reasoning;  
(3) \textbf{DynaMath}~\cite{DynaMath}: dynamic visual math reasoning benchmark for evaluating robustness of vision-language models under challenging and shifting distributions.
(4) \textbf{AI2D}~\cite{ai2d}: single-image diagram understanding;  
(5) \textbf{InfoVQA}~\cite{docvqa}: single-image information graphics reasoning;  
(6) \textbf{TextVQA}~\cite{textvqa}: single-image text understanding via OCR;  
(7) \textbf{MMBench}~\cite{mmbench}: single-image multimodal QA with a mix of perception, reasoning, and common-sense tasks.  

This selection spans single-image, multi-image, and relatively text-heavy scenarios, enabling evaluation of model accuracy across a variety of cross-modal reasoning challenges and ensuring that our load balancing and precision strategies are tested under diverse multimodal conditions.

\textbf{Testbed.} 
Our target deployment environment is large-scale server-grade GPU clusters based on the Blackwell architecture (e.g., multi-node B100 or B200 systems), where large expert parallelism (EP) exacerbates device-level load imbalance.
Due to hardware availability, all experiments are conducted on a single node with 8 NVIDIA RTX 5090 GPUs (64\,GB/s).

As RTX 5090 is a consumer-grade GPU with limited interconnect bandwidth, MoE inference end-to-end performance in this setting is heavily communication-bound compared to server-class GPUs. 
To better emulate server deployment, we measure the communication latency on NVIDIA H20 GPUs using NVLink (4 TB/s) and replace the communication component in the 5090 end-to-end measurements with these high-bandwidth values.
This approach preserves the computation characteristics of the 5090 setup while providing an estimate of e2e performance under high-bandwidth interconnects. 
The software and hardware configurations are summarized in \Cref{tab:hardware_config}.
\begin{table}[h!]
\centering
\caption{Software and hardware configuration for experiments.}
\label{tab:hardware_config}
\begin{tabular}{l l}
\toprule
\textbf{Component} & \textbf{Configuration} \\
\midrule
vLLM & v0.13.0 \\
Lmms-eval & v0.5 \\
Sample Params & topp=0.8,topk=20,temperature=0.7 \\
EPLB & 8 redundant experts, window size=100, interval=100 \\
Python             & 3.12  \\
PyTorch            & 2.8.0 \\
CUDA               & 12.8 \\
OS & Ubuntu 22.04 \\
GPU                & 8 $\times$ NVIDIA RTX 5090 (32\,GB each) \\
Interconnect       & Gen5 PCIe, 64\,GB/s per GPU \\
CPU                & 100 vCPU Intel Xeon Platinum 8470Q \\
\bottomrule
\end{tabular}
\end{table}

\textbf{Quantization Details.}
We adopt the NVFP4 (W4A4) scheme, where weights and activations are represented in FP4 (E2M1), and scaling factors are stored in FP8 (E4M3). Quantization is performed in a per-group manner with a group size of 16.
For each group, we apply symmetric min-max quantization. Specifically, we compute the dynamic range as 
 and derive the local scale by normalizing with the maximum representable FP4 value (6.0). A global scale factor is further applied to align magnitudes across groups or tensors. The resulting scale is then quantized into FP8 and clamped to its valid range, ensuring numerical stability. Each value is quantized by dividing by the scale and mapping to the nearest FP4 representable value. During inference, matrix multiplication is performed directly on FP4 operands.

\section{Routing Statistics}
\label{sec:append_routing}
\textbf{Device-level load imbalanceness.}
\Cref{fig:routing_layers2} shows device-level load imbalance (1× to 3×) for Kimi-VL under EP=8, measured on the MMMU dataset.


\begin{figure*}[b!]
    \centering
    \foreach \i in {12,...,21} {
        \begin{subfigure}{0.48\linewidth}
            \includegraphics[width=\linewidth]{Figures/append/rank_load_imbalance/layer_\i_rank_imbalance.pdf}
            \caption{Layer \i}
        \end{subfigure}
        \ifodd\i \hfill \fi
    }
    \caption{Dynamics of device-level load imbalance across MoE layers 12-21.}
    \label{fig:routing_layers2}
\end{figure*}

\textbf{Dynamics of load distribution across devices, experts, and modalities.}  
\Cref{fig:hybrid_mmmu_layers2} shows several representative MoE layers, illustrating how the load distribution of each device and the most selected expert (top-1 hot expert) evolve over iterations. The plots also indicate the proportion of different modality tokens, highlighting workload skew, modality imbalance, and the persistence of hotspots during inference.



\begin{figure*}[t!]
    \centering
    \foreach \i in {12,...,19} {
        \begin{subfigure}{0.48\linewidth}
            \includegraphics[width=\linewidth]{Figures/append/hybrid_mmmu_iter_1/layer_\i_expert_rank_load_broken.pdf}
            \caption{Layer \i}
        \end{subfigure}
        \ifodd\i \hfill \fi
    }
    \caption{Overloaded devices and experts across iterations for layer 12-19, with token proportions for different modalities.}
    \label{fig:hybrid_mmmu_layers2}
\end{figure*}

\clearpage

\section{Batching pattern}
\label{sec:append_batching}

\textbf{Mixed prefill and decode scheduling.}
Mixing prefill and decode tokens within a batch naturally arises from the continuous batching mechanism in vLLM under a colocated deployment setting. In this setup, prefill and decode requests are not separated into distinct instances. Instead, the scheduler dynamically forms batches from incoming requests at each step. Moreover, chunked prefill further splits long prefill requests into smaller segments, allowing them to be scheduled together with decode requests within the same batch. This design improves overall utilization and avoids decode starvation. Under this setting, we define an iteration as one continuous batching step, corresponding to a single model forward pass.

\textbf{Comparison with disaggregated deployment.}
In contrast, prefill/decode-disaggregated deployment assigns prefill and decode workloads to different instances, typically requiring duplicated resources. Under such settings, ReaLB mainly applies to prefill workers, where the majority of computation and routing decisions occur.

\begin{figure*}[tbh]
    \centering
    \includegraphics[width=0.6\linewidth]{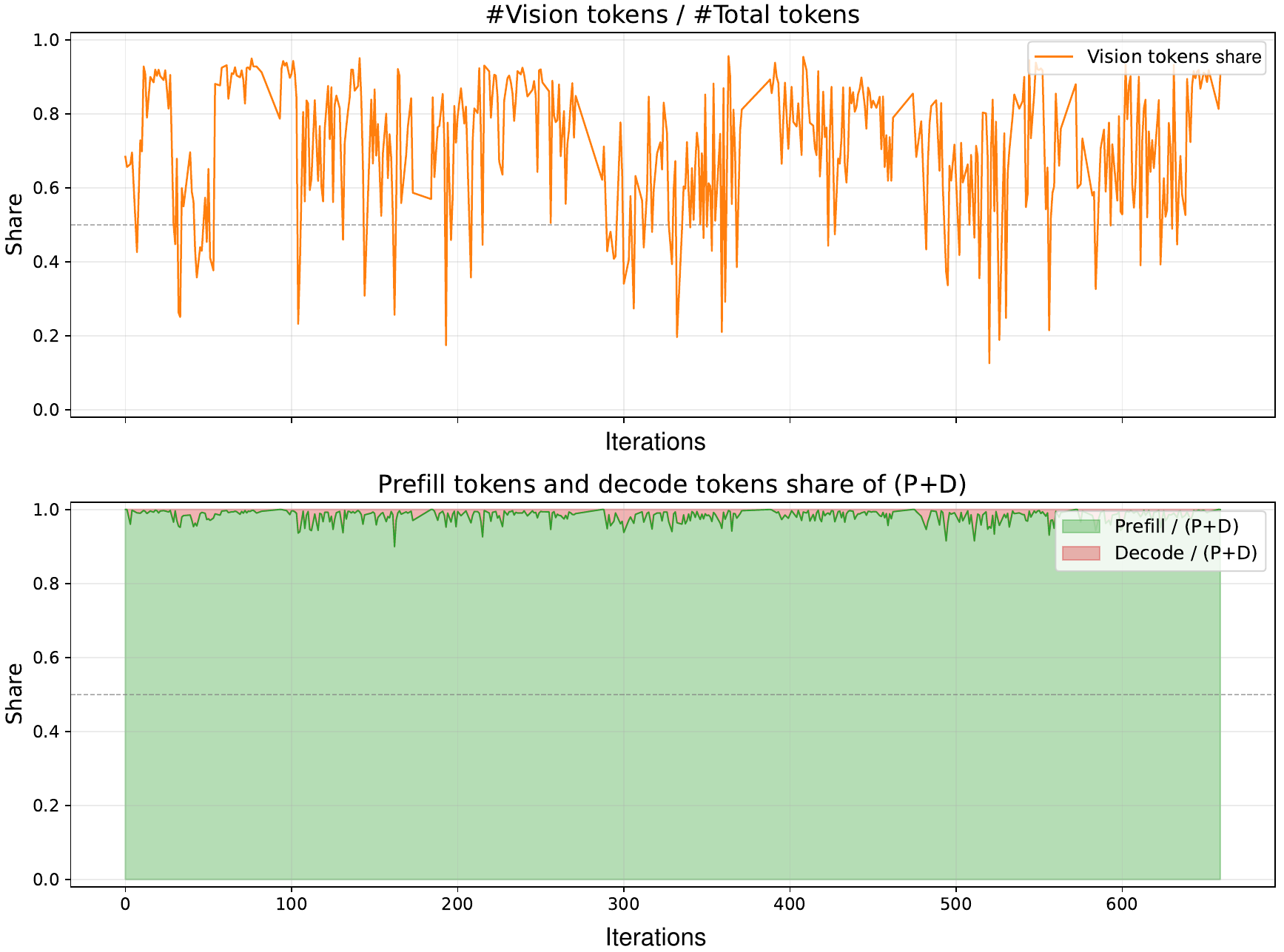}
    \caption{Token composition in continuous batching.}
    \label{fig:batch_token_ratio}
\end{figure*}

\textbf{Prefill-dominated batches.}
Although batches may contain mixed prefill and decode tokens, they are strongly dominated by prefill tokens in our workload. With \texttt{max-num-seqs=12}, a mixed batch contains at most 11 decode tokens (one per request), while prefill tokens occupy the remaining capacity. In practice, decode tokens account for less than 10\% of the total tokens per batch, as shown in \Cref{fig:batch_token_ratio}. As a result, under our colocated deployment (used due to limited GPU resources), the observed routing dynamics are largely dominated by prefill tokens, providing a close approximation to a prefill-only setting without explicitly separating prefill and decode execution.

\begin{table}[hbt]
\centering
\caption{Speedup of ReaLB in the prefill-only setting.}
\label{tab:prefill}
\begin{tabular}{c|ccc}
\toprule
Model &
MMMU & MathVista & DynaMath \\
\midrule

Kimi-VL
& 1.35$\times$ & 1.33$\times$ & 1.34$\times$  \\

\midrule

Qwen-VL
& 1.12$\times$ & 1.12$\times$ & 1.10$\times$  \\

\bottomrule
\end{tabular}

\vspace{-0.8em}
\end{table}
 


\section{AIMD-based Adaptive Control Analysis}
\label{app:aimd}

We analyze the dynamics of the AIMD-based adaptive control on real-world vision-language tasks.
The congestion threshold $\tau$ is set to 1.5, above which the system is considered to be under severe imbalance (highlighted by the red background in the figures).

\begin{figure}[thb]
    \centering

    \begin{subfigure}[t]{0.45\linewidth}
        \centering
        \includegraphics[width=\linewidth]{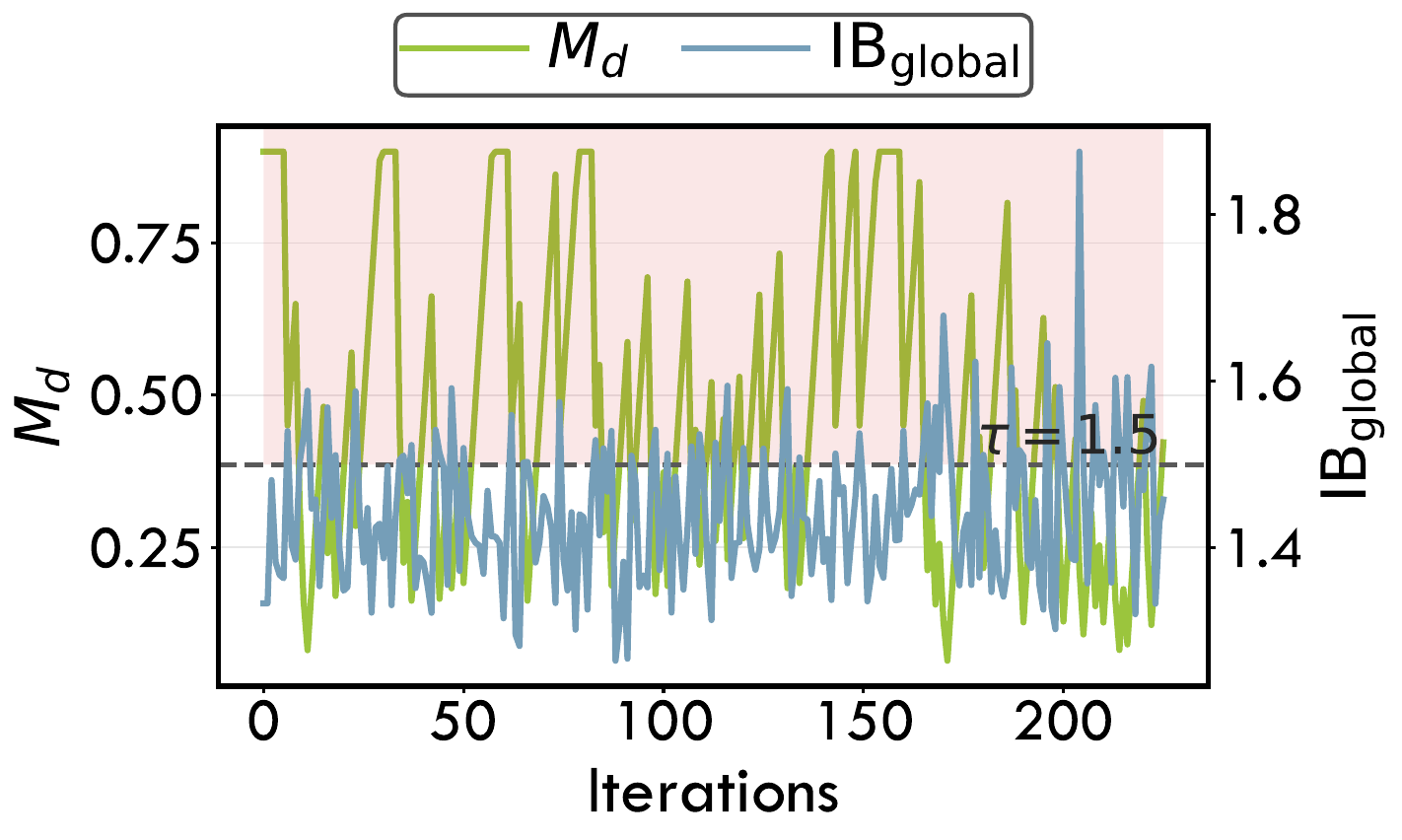}
        \caption{Kimi-VL.}
    \end{subfigure}
    \hfill
    \begin{subfigure}[t]{0.45\linewidth}
        \centering
        \includegraphics[width=\linewidth]{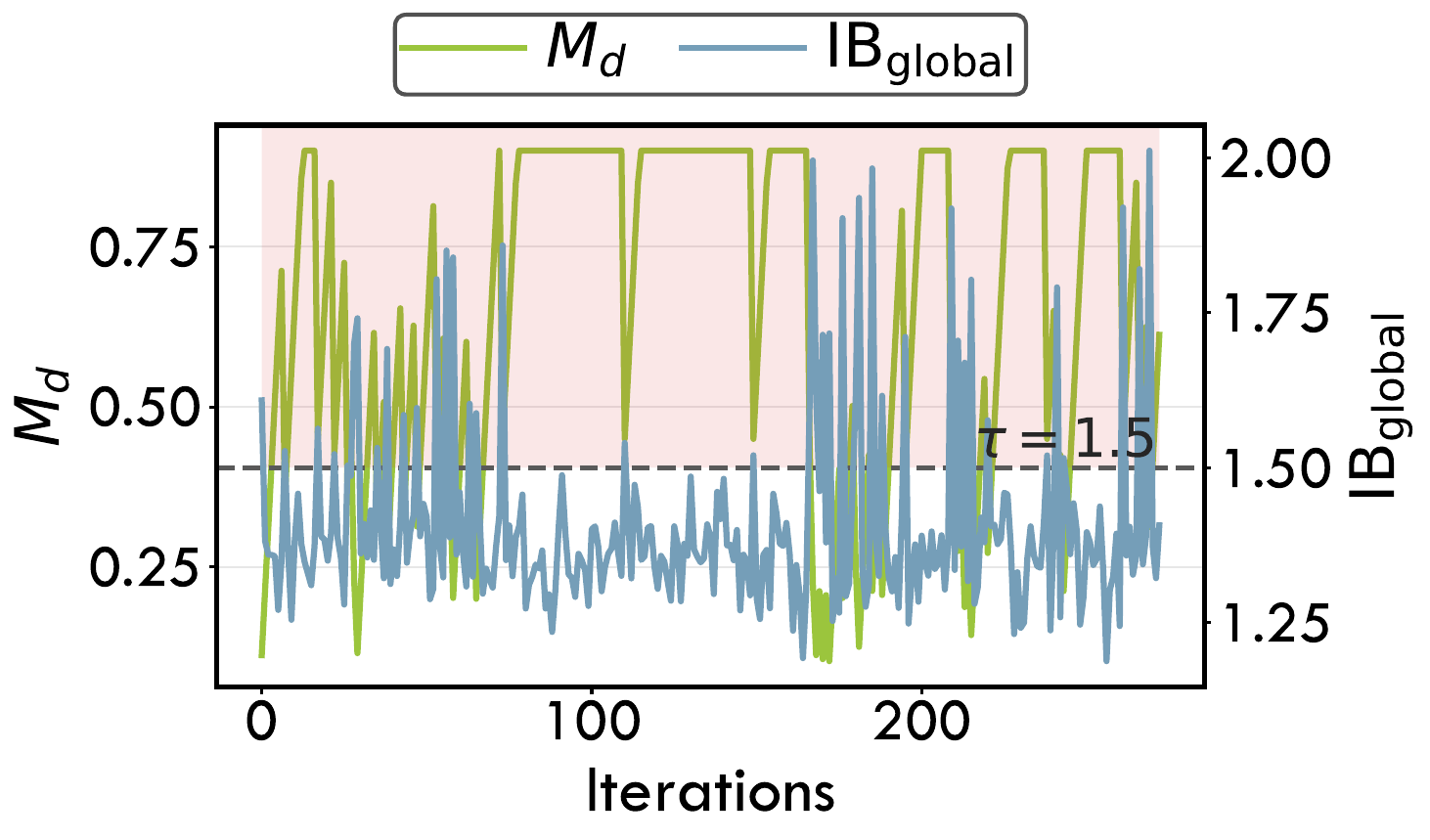}
        \caption{Qwen3-VL.}
    \end{subfigure}

    \caption{Evolution of the control parameter $M_d$ under AIMD-based adaptive control on DynaMath. 
    The red background indicates time steps where the global imbalance exceeds the congestion threshold.}
    \label{fig:aimd}
\end{figure}

\Cref{fig:aimd} illustrates the evolution of the control parameter $M_d$ over time for two representative multimodal MoE models.
When the global imbalance exceeds the threshold, $M_d$ is rapidly decreased, enabling more aggressive acceleration for overloaded devices with high concentrations of visual tokens, thereby improving execution efficiency.
In contrast, when the system operates below the congestion threshold, $M_d$ gradually increases, relaxing the constraint to preserve model accuracy.

These results show that the AIMD-based controller can effectively adapt to dynamic imbalance conditions, achieving a stable balance between efficiency and accuracy.

\section{Additional Efficiency Results}
\label{app:efficiency_more}

\paragraph{End-to-end Latency Breakdown.}
We further report end-to-end latency breakdown on two additional benchmarks, MMMU and MathVista, for Kimi-VL and Qwen3-VL, which are not included in the main paper due to space constraints.

\begin{figure*}[thb]
    \centering

    \begin{subfigure}[b]{0.45\textwidth}
        \centering
        \includegraphics[width=\linewidth]{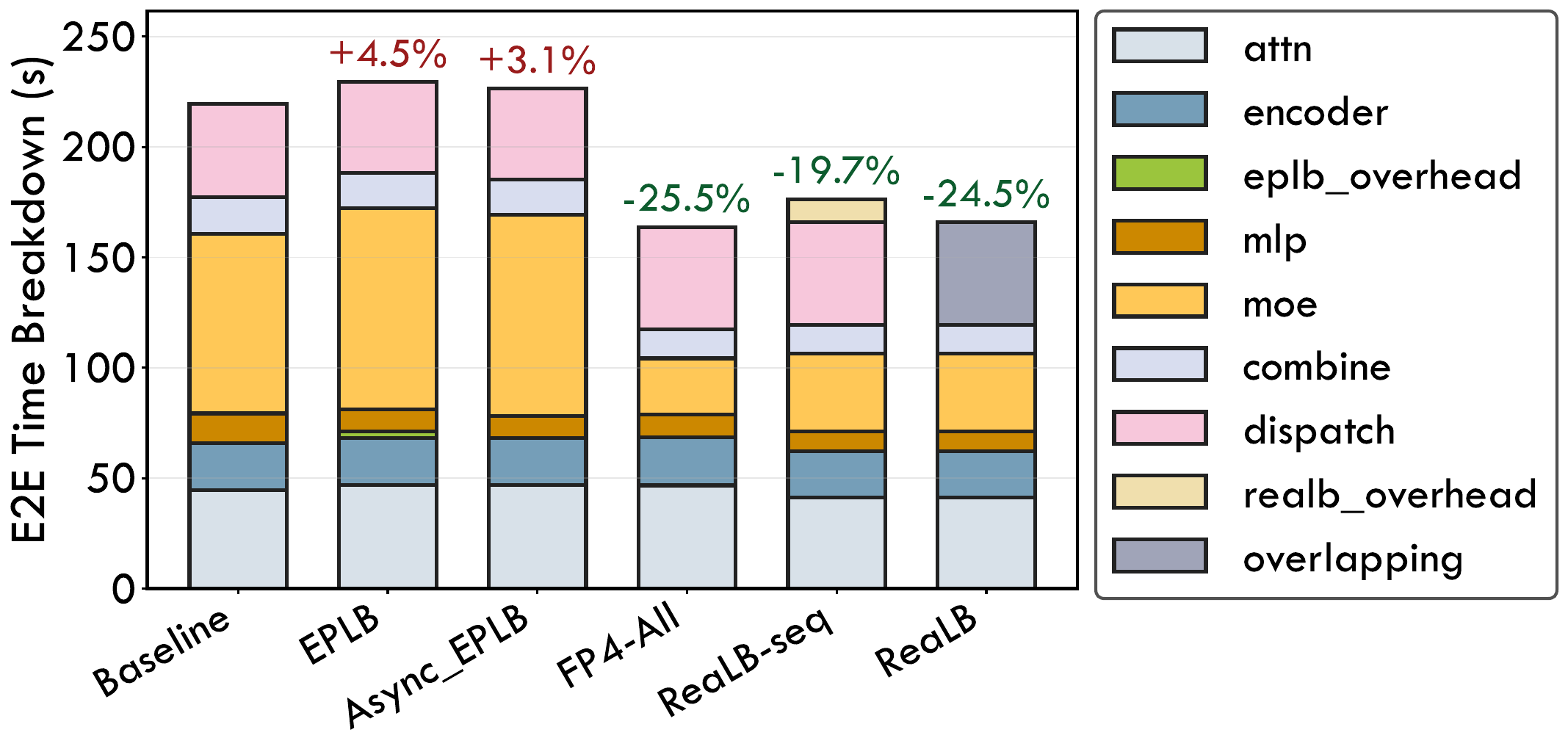}
        \caption{Kimi-VL, MMMU.}
    \end{subfigure}
    \hspace{0.05\textwidth}
    \begin{subfigure}[b]{0.45\textwidth}
        \centering
        \includegraphics[width=\linewidth]{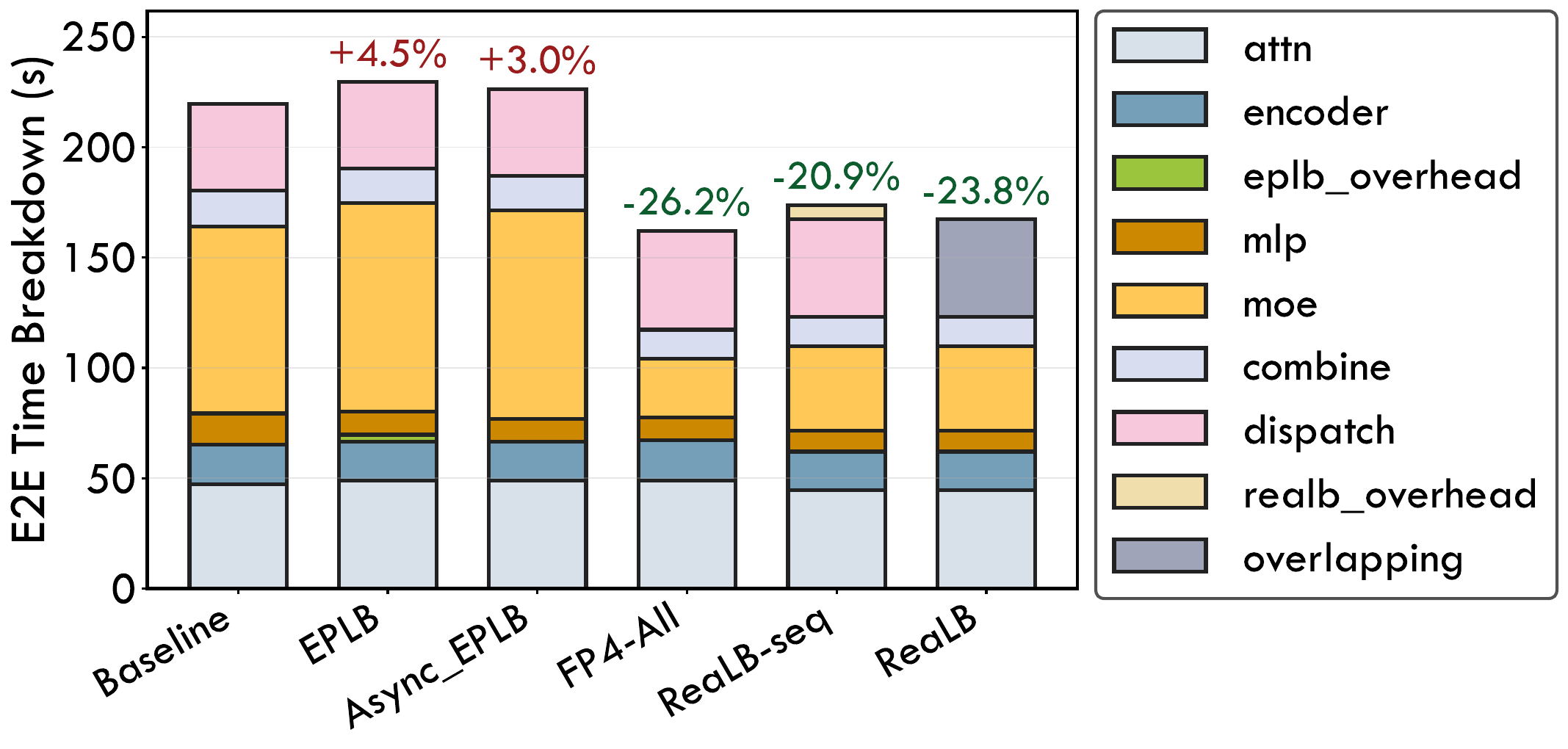}
        \caption{Kimi-VL, MathVista.}
    \end{subfigure}

    \begin{subfigure}[b]{0.45\textwidth}
        \centering
        \includegraphics[width=\linewidth]{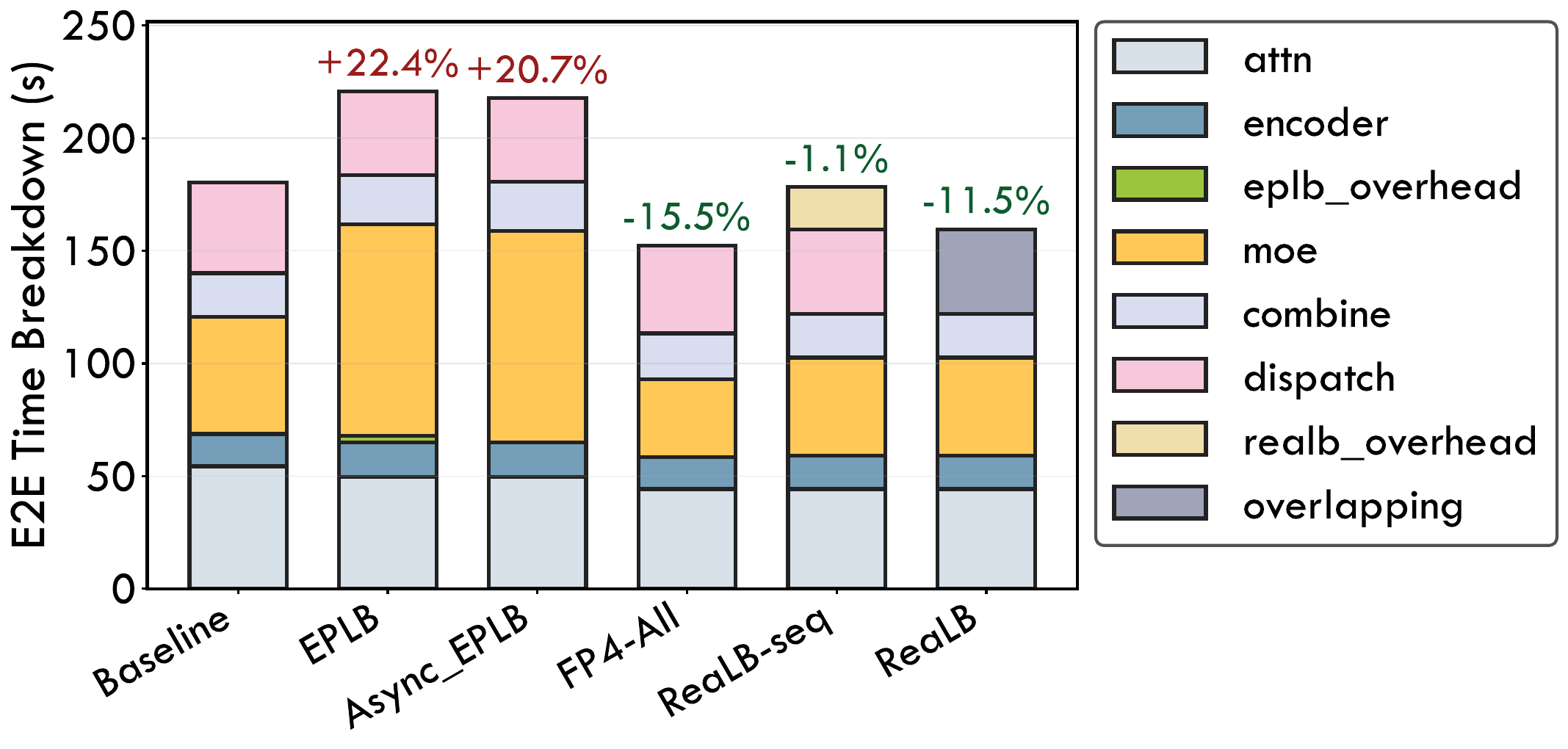}
        \caption{Qwen3-VL, MMMU.}
    \end{subfigure}
    \hspace{0.05\textwidth}
    \begin{subfigure}[b]{0.45\textwidth}
        \centering
        \includegraphics[width=\linewidth]{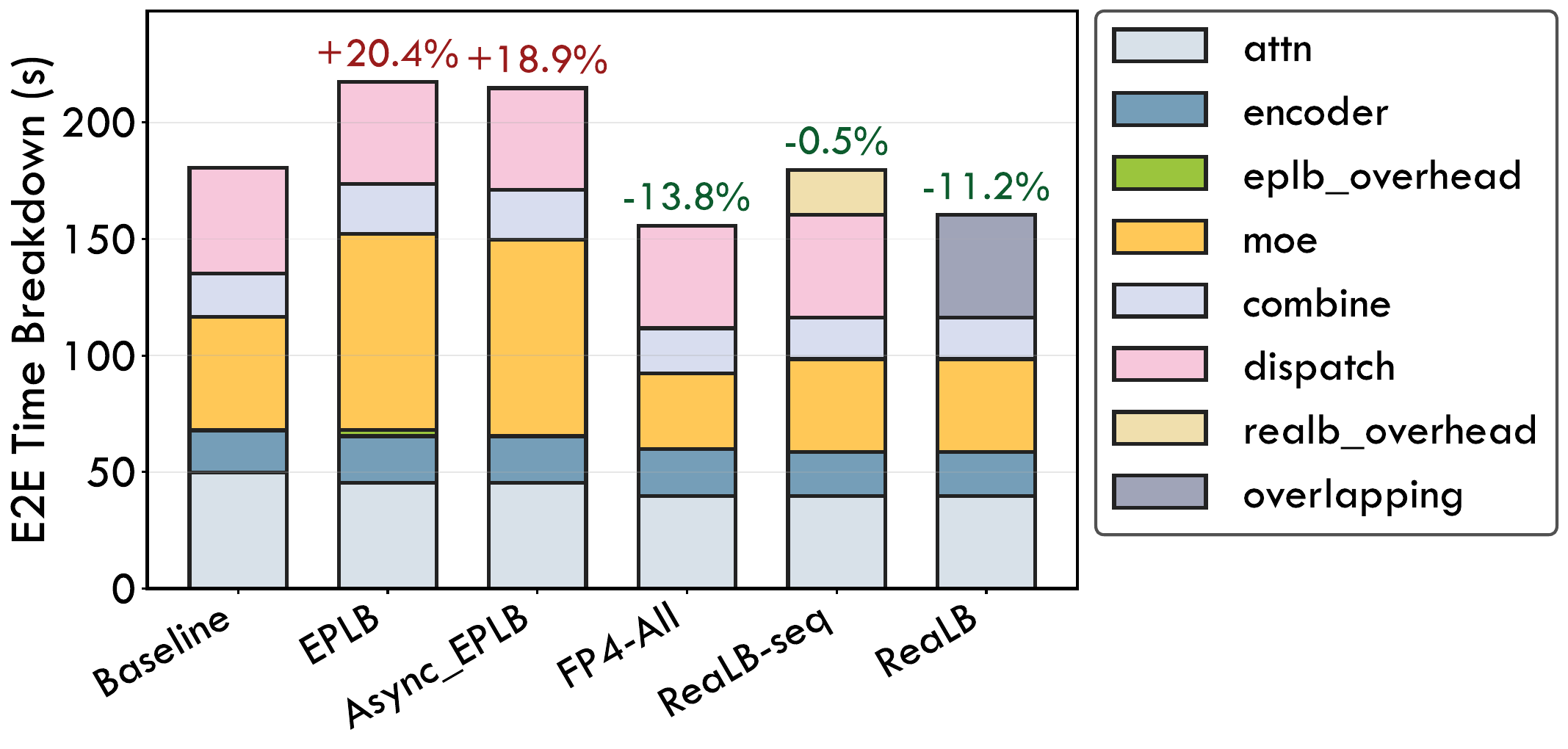}
        \caption{Qwen3-VL, MathVista.}
    \end{subfigure}

    \caption{End-to-end latency breakdown on additional benchmarks.}
    \label{fig:efficiency}
\end{figure*}





\end{document}